# POISSON—LIE GROUP OF PSEUDODIFFERENTIAL SYMBOLS

BORIS KHESIN AND ILYA ZAKHAREVICH

August 1993   Printed: October 20, 1993

ABSTRACT. We introduce a Lie bialgebra structure on the central extension of the Lie algebra of differential operators on the line and the circle (with scalar or matrix coefficients). This defines a Poisson—Lie structure on the dual group of pseudodifferential symbols of an arbitrary real (or complex) order. We show that the usual (second) Benney, $GL_n$-KdV (or $GL_n$-Adler—Gelfand—Dickey) and KP Poisson structures are naturally realized as restrictions of this Poisson structure to submanifolds of this "universal" Poisson—Lie group. Moreover, the reduced ($= SL_n$) versions of these manifolds (or $W_n$-algebras in physical terminology) can be viewed as certain subspaces of the quotient of this Poisson—Lie group by the dressing action of the group of functions on the circle (or as a result of a Poisson reduction). Finally we define an infinite set of commuting functions on this Poisson—Lie group that give the standard families of Hamiltonians when restricted to the submanifolds mentioned above. The Poisson structure and Hamiltonians on the whole group interpolate between the Poisson structures and Hamiltonians of Benney, KP and KdV flows. We also discuss the geometrical meaning of $W_\infty$ as a limit of Poisson algebras $W_\varepsilon$ when $\varepsilon \to 0$.

## CONTENTS

## 0. Introduction

Being a fashionable exercise during the last twenty-odd years, the theory of integrable systems looks now like a patched quilt. Numerous teams carried out profound investigations of their patches, and if one restricts attention to a particular point of view, the picture is usually transparent now and looks rather complete. Problems begin to arise whenever we remember that a particular phenomenon in integrable systems can be described from two (or more) different points of view. Usually both explanations are satisfactory, but often there is no clear way to relate them one to another.

The result is a sophisticated "manifold" where the "transition functions" are no less important (and much more difficult to establish) than "local charts" themselves. In this paper we describe a "transition function" from the "quantum groups" chart (more precise, its quasiclassical limit dealing with Poisson—Lie groups) to the chart dealing with geometry of differential operators. The pivotal point in this transition is the equipment of the Kac—Peterson central extension of the Lie algebra of differential operators with a Lie bialgebra structure.

Consideration of the corresponding Manin—Drinfeld double introduces into the cast the Lie algebra of pseudodifferential symbols. The algebras of differential and pseudodifferential symbols play an important rôle in the theory of integrable hierarchies (of types KdV or KP) [32] and in conformal field theory (with $W_n-$ and $W_\infty$-symmetry in physics-speech). There exist several approaches to these algebras describing their central extensions, pairs of compatible Poisson brackets (so called Gelfand—Dickey brackets), connections between KP and $n$-KdV equations. In this paper we describe these structures from the unified viewpoint of Poisson—Lie group of pseudodifferential symbols and of a certain extension of this group. There are two different ways to introduce this extension, that correspond to two mutually dual Poisson—Lie "groups".

The first approach begins with introduction of a new element $\log D$. It turns out that the (Lie) algebra of integral symbols $\sum_{i<0} u_i D^i$ (i.e., the Volterra algebra) extended by the new symbol $\log D$ can be integrated to a remarkable Lie group. This group consists of classical pseudodifferential symbols of an arbitrary real (or complex) order, and carries a natural Poisson—Lie structure.

The second approach begins with the Lie algebra of differential operators $\sum_{i \geq 0} u_i D^i$. It turns out that the introduced by Kac and Peterson [15] 1-dimensional central extension of this Lie algebra also carries a natural Poisson—Lie (=Lie bialgebra) structure. Although this Lie algebra cannot be integrated to a Lie group, it is the dual Lie algebra of the mentioned above Poisson—Lie group of extended integral symbols.

Our main object of study is this Poisson—Lie structure on the Lie group of classical pseudodifferential operators. We verify that this structure restricted to the submanifold of differential operators coincides with the second Gelfand—Dickey Poisson structure, and restricted to a KP-phase space coincides with the quadratic KP structure. Moreover, when extended on the whole group, KP and KdV Hamiltonians correspond to the center

$$\mathrm{Cent}\left(\mathrm{Func}\,(\mathrm{Group})^{\mathrm{Ad}_{\mathrm{Group}}}\right)$$

of the Poisson algebra of invariant functions on this group.

This puts the schemes for KP and KdV hierarchies into the framework of the Poisson—Lie geometry on the group of pseudodifferential symbols. The analogous properties are valid for a submanifold of operators of order 0, the corresponding Poisson structure is the second Benney structure. As corollaries of our construction we obtain the Poisson properties of the multiplication of differential operators, of the Miura mapping (or Kupershmidt—Wilson theorem [21]), and generalize the Radul theorem [29] on the Poisson action of the Lie algebra of all differential operators on the set of fixed order differential operators (with the leading coefficient 1) equipped with the second Gelfand—Dickey Poisson structure. Due to all this we can conceive the Poisson structure on the above group as a structure that interpolates between (second) Benney, KP and $\mathrm{GL}_n$-Gelfand—Dickey structures.

**0.1. Structure of the paper.** The paper is organized as follows. In Section 1 we recall the basic notions of the theory of Poisson—Lie groups and of Lie bialgebras, as well as the relation of one to the other. In Section 2 we define the Lie algebra of pseudodifferential symbols and discuss how to add to it a new symbol $\log D$. Here we also discuss an extension of the Kac—Peterson cocycle to the Lie algebra of pseudodifferential operators. In Section 3 we integrate the Lie algebra consisting of integral symbols together with the symbol $\log D$ to a Lie group. In Section 4 we introduce a Lie bialgebra structure on the preceding Lie algebra. To do this we construct the Manin double of the would-be Lie bialgebra. In fact we fuse two described above approaches and add simultaneously the symbol $\log D$ and the cocycle to pseudodifferential operators. As the results of Section 1 show, this defines a Poisson—Lie structure on the group from Section 3. This brings to an end the first logical part of the paper, the part devoted to definition of this Poisson—Lie structure.

Beginning from Section 5 we study the resulting Poisson structure. In fact this section is just a warm-up: the Poisson—Lie structure on the Volterra group of classical symbols of order 0 turns out to be the known among specialists so called second Benney structure. In Section 6 we generalize this description to the ambient group of arbitrary order pseudodifferential symbols. The Poisson bracket written in global coordinates is the generalized Gelfand—Dickey—KP structure. In Section 7 we use this to give one-line proofs of some facts about (second) $n$-KdV- and KP-structures that usually demand long-winged calculations. It is the Jacobi identity for the bracket and the facts that the multiplication of differential operators, and, in particular, Miura transform are Poisson mappings.

In Section 8 we investigate the relation of extended (= GL) KP structure and the usual (= SL) KP structure. We show that in the Poisson—Lie language it is a Poisson reduction with respect to the (dressing=adjoint) action of Lie subalgebra of functions on the circle. (After taking the Kac—Peterson central extension of differential operators this subalgebra ceases to be commutative and becomes a Heisenberg algebra).

In Section 9 we provide a Poisson—Lie framework for the Hamiltonians on various integrable systems contained inside our Poisson—Lie group. In particular, we interpolate the KP and $n$-KdV hierarchies into a unified hierarchy on this group. One of byproducts is a purely algebraic description of these (KP type) Hamiltonians that is valid for any Poisson—Lie group. In Section 11 we discuss the quantization of this notion.

In the same Section 9 we provide an expression of the Poisson bracket on a Poisson—Lie group in exponential coordinates that is of independent interest. Note that this expression makes it possible to reduce a usual two-step process of construction of the Poisson—Lie structure to a new one-step process: instead of construction of the Manin—Drinfeld double basing on the corresponding $r$-matrix, and then description of the Poisson—Lie group as a Poisson submanifold in this double, we can construct the Poisson structure in a neighborhood of the unity using Formula (9.1).

In Section 10 we introduce two more Poisson—Lie groups. Both these groups are 1 dimension bigger than our main object of study. The Manin—Drinfeld doubles of these groups coincide, and this double is in some sense *a universal extension* of the Volterra group. This corresponds to a fact that Lie algebra of pseudodifferential operators allows two nonhomotopic cocycles. At last, in Section 11 we sketch what are impacts of results obtained here on a would-be quantization of the KP structure. First of all, we describe analogues of KP Hamiltonians for any Hopf algebras, in particular, for such a quantization. Secondly, we describe a relation between $W_\infty$-algebra and this quantization. We show that $W_\infty$-algebra is a "linearization" of this quantization up to a change of the order of evaluation of two limits. Here "linearization" is a vague resemblance of the relation between a Lie group and a Lie algebra (up to a fact that Hopf algebras are intrinsically nonlocal).

**0.2. Motivations and different viewpoints.** In a sense we describe in this paper a fusion of two ways to generalize the KdV equation, considered as an Euler—Arnold equation on the dual space to the Virasoro algebra [2, 13, 27]. This dual space carries a linear Poisson structure, what is "the same" as a Lie algebra structure on the dual space. Two principal ways to generalize the KdV equation comes from generalizations of structures on these two mutually dual vector spaces. We can diagram these ways as

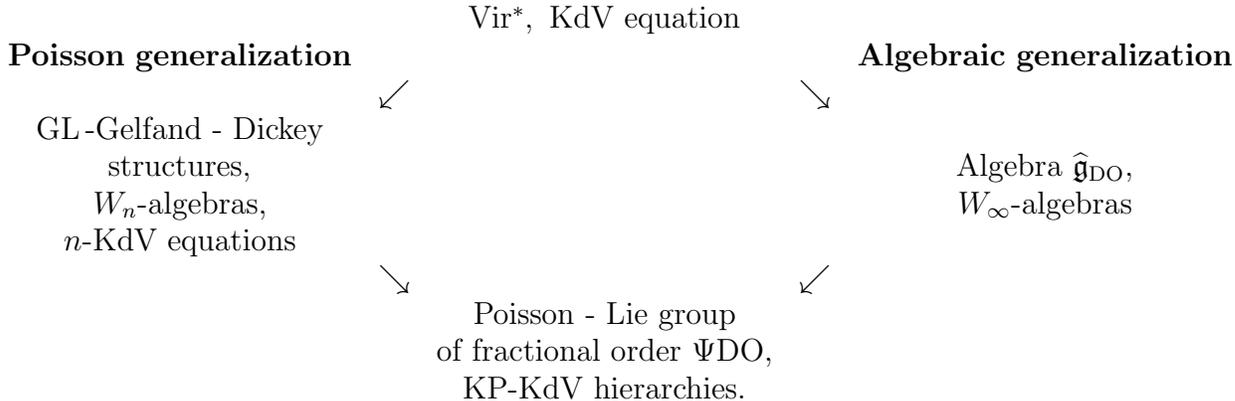

The first approach (earmarked by physicists as "classical $W_n$-algebras") comes from consideration of quadratic Poisson structures instead of linear ones. It leads to the (quadratic) Gelfand—Dickey Poisson structure on coefficients of differential operators of order $n$. The second approach (earmarked as $W_\infty$-algebras) leads to consideration of the bigger Lie algebra, namely the Lie algebra of differential operators of an arbitrary order (instead of vector fields).

Here we show that the recognized the long time ago correspondence between quadratic brackets and Poisson—Lie groups allows one to fuse these two approaches together. The only missing step was the Lie bialgebra structure on the central extension of Lie algebra of differential operators and integration of the dual Lie algebra to a Lie group.

There is another point of view on the results obtained here. Here we begin with definitions of some extensions of Poisson—Lie groups and show that the second Gelfand—Dickey Poisson structure is a submanifold of one of this group. We conclude that the dressing action determines the (Radul) action of the Lie algebra differential operators on a Poisson manifold consisting of differential operators (of fixed order). On the other hand, we could begin with the Radul action as with the base object. This action can be considered as an action of a Poisson—Lie algebra (=Lie bialgebra) on an abstract Poisson manifold.

Let us consider the corresponding momentum mapping. The usual construction of the momentum mapping can be carried out in the case when the action of a Lie algebra preserves a Poisson structure. In this case we can construct (locally) a central extension of the Lie algebra and a mapping from the manifold to the dual space to this central extension. However, a Poisson—Lie action preserves the Poisson structure only if the Poisson structure on the Lie algebra vanishes (give or take a couple of counterexamples).

It is interesting to try to generalize the construction of the momentum mapping to the general case of an arbitrary Poisson—Lie action. Though there seems to be no such generalization, in some particular cases one *can* construct a Poisson—Lie central extension and a mapping from an open subset of the manifold to the *dual group* to this central extension (the "prequantized" momentum mapping).

Thus the results obtained here can be interpreted as constructions of this central extension, of the corresponding dual group and of the inclusion of the manifold in question into this dual group in the case of Radul action. In the paper [38] one of the authors shows that the second Gelfand—Dickey Poisson structure is a structure on an open subset of some Poisson—Lie Grassmannian, and the Radul action is the restriction to the Lie subalgebra of differential operators of the natural action of $\mathfrak{gl}$ on the Grassmannian. Therefore we can view the results of this paper as a description of the "prequantized" momentum mapping for the Grassmannian.

Certain main results of this paper have been reported in [17]. For a discussion of a Lie-Poisson structure on pseudodifferential operators see also in a recent preprint [8], some questions related to the family of integrable KdV-type hierarchies were considered in the paper [11].

**Acknowledgments.** We are pleased to thank V. Arnold, J. Bernstein, E. Frenkel, I. Frenkel, K. Gawedski, I. M. Gelfand, M. Gromov, S. Khoroshkin, T. Khovanova, A. Radul, T. Ratiu, N. Reshetikhin, C. Roger, V. Rubtsov, G. Segal, M. Semenov-Tian-Shansky, V. Serganova, A. Weinstein and G. Zuckerman for fruitful discussions. The first author expresses his gratitude to the Institut des Hautes Etudes Scientifiques in Bures-sur-Yvette, to the University Paris-7, and especially to Professors M. Chaperon and C. Roger for their kind hospitality. The paper would not see the light of day without a cordial support from V. Serganova, to whom the second author is limitlessly indebted. This work was partially supported by the NSF grant DMS-9307086.

## 1. Lie bialgebras and Poisson—Lie Groups

**1.1. Basic definitions.** In this section, following [26] and [34], we recall the reader some necessary facts about Lie bialgebras and Poisson—Lie structures.

**Definition 1.1.** A Poisson manifold is a manifold with a *Poisson bracket* on the set of functions on this manifold, i.e., with a skewsymmetric operation on functions that satisfies the Leibniz

$$\{f, gh\} = g\{f, h\} + \{f, g\}h$$

and the Jacobi identities.

**Definition 1.2.** A Poisson Lie group $(G, \eta)$ is a Lie group equipped with a Poisson structure $\eta$ such that the multiplication $G \times G \to G$ and the inversion mapping

$$i \colon G \to G^-, \qquad g \mapsto g^{-1}$$

are Poisson maps, where $G \times G$ carries the product Poisson structure, and $G^-$ is the group $G$ equipped with the opposite Poisson structure.

This property can be formulated in terms of the corresponding Lie algebra $\mathfrak{G}$. Consider a pair of vanishing at $e$ functions on $G$. The above property of the inversion mapping $i$ shows that the Poisson bracket of this pair also vanishes at $e$. Moreover, the linear part of this bracket at $e$ is uniquely determined by the linear parts of the original functions. This defines a Lie algebra operation on $\mathfrak{G}^*$. A simple check shows that it satisfies the restriction from the following

**Definition 1.3.** A pair $(\mathfrak{G}, \mathfrak{G}^*)$ of a Lie algebra $\mathfrak{G}$ and its dual space $\mathfrak{G}^*$ forms a Lie bialgebra if $\mathfrak{G}^*$ is equipped with a Lie algebra structure such that the map $\mathfrak{G} \to \mathfrak{G} \wedge \mathfrak{G}$ dual to the Lie bracket map $\mathfrak{G}^* \wedge \mathfrak{G}^* \to \mathfrak{G}^*$ on $\mathfrak{G}^*$ is a 1-cocycle on $\mathfrak{G}$ relative to the adjoint representation of $\mathfrak{G}$ on $\mathfrak{G} \wedge \mathfrak{G}$.

**Theorem 1.4 (see [26]).** *For any connected and simply connected group $G$ with a Lie algebra $\mathfrak{G}$ there is a one-to-one correspondence between Lie bialgebra structures on $(\mathfrak{G}, \mathfrak{G}^*)$ and Poisson—Lie structures $\eta$ on $G$. This correspondence sends a Poisson—Lie group $(G, \eta)$ into the tangent to $(G, \eta)$ Lie bialgebra $(\mathfrak{G}, \mathfrak{G}^*)$.*

*Remark* 1.5. We discuss the details of this construction in Section 5.

Equivalently, one can describe the structure on a Lie algebra of a Poisson—Lie group using a Manin triple $\left( \bar{\mathfrak{G}}, \mathfrak{G}_+, \mathfrak{G}_- \right)$ (for which $\mathfrak{G}_+ = \mathfrak{G}$, $\mathfrak{G}_- = \mathfrak{G}^*$ and $\bar{\mathfrak{G}} = \mathfrak{G} \oplus \mathfrak{G}^*$):

**Definition 1.6.** A Manin triple consists of a triple of Lie algebras $\left( \bar{\mathfrak{G}}, \mathfrak{G}_+, \mathfrak{G}_- \right)$ and a nondegenerate invariant symmetric inner product $\langle , \rangle$ on $\bar{\mathfrak{G}}$ such that

  (1) $\bar{\mathfrak{G}} = \mathfrak{G}_+ \oplus \mathfrak{G}_-$ as a vector space;
  (2) both $\mathfrak{G}_+$ and $\mathfrak{G}_-$ are Lie subalgebras of $\bar{\mathfrak{G}}$;
  (3) both $\mathfrak{G}_+$ and $\mathfrak{G}_-$ are isotropic with respect to the inner product $\langle , \rangle$.

**Theorem 1.7 (see [34]).** *Consider a Manin triple $\left( \bar{\mathfrak{G}}, \mathfrak{G}_+, \mathfrak{G}_- \right)$. Then $\mathfrak{G}_+$ is naturally dual to $\mathfrak{G}_-$, and the pair $\left( \mathfrak{G}_-, \mathfrak{G}_-^* = \mathfrak{G}_+ \right)$ is a Lie bialgebra. Moreover, for any Lie bialgebra $(\mathfrak{G}, \mathfrak{G}^*)$ one can find a unique Lie algebra structure on $\overline{\mathfrak{G}} = \mathfrak{G} \oplus \mathfrak{G}^*$* such that the triple $\left( \overline{\mathfrak{G}}, \mathfrak{G}, \mathfrak{G}^* \right)$ is a Manin triple with respect to the natural pairing on $\overline{\mathfrak{G}}$:

$$((X_1, \alpha_1), (X_2, \alpha_2)) = \langle X_1, \alpha_2 \rangle + \langle X_2, \alpha_1 \rangle,$$

and the corresponding Lie bialgebra structure on $(\mathfrak{G}, \mathfrak{G}^*)$ is the given one.

*Remark* 1.8. The Lie bracket on $\bar{\mathfrak{G}} = \mathfrak{G}_+ \oplus \mathfrak{G}_-$ (here $\mathfrak{G}_+ = \mathfrak{G}$, $\mathfrak{G}_- = \mathfrak{G}^*$) is given by

$$[x_1 + \alpha_1, x_2 + \alpha_2]_{\bar{\mathfrak{G}}} = [x_1, x_2]_{\mathfrak{G}} - \mathrm{ad}^*_{\alpha_2} x_1 + \mathrm{ad}^*_{\alpha_1} x_2 + [\alpha_1, \alpha_2]_{\mathfrak{G}^*} + \mathrm{ad}^*_{x_1} \alpha_2 - \mathrm{ad}^*_{x_2} \alpha_1$$

where $x_1, x_2 \in \mathfrak{G}$, $\alpha_1, \alpha_2 \in \mathfrak{G}^*$ (see [26]).

## 1.2. Poisson—Lie subgroups and Poisson—Lie actions.

The usual notions of a Lie subalgebra and of a Lie subgroup are almost useless in most contexts of the Poisson—Lie groups theory. A good substitute is a notion of a Poisson—Lie subalgebra and a Poisson—Lie subgroup.

**Definition 1.9.** A submanifold $N$ of a Poisson manifold $M$ is a *Poisson submanifold* if the restriction to $N$ of the Poisson bracket of two functions on $M$ is uniquely determined by the restrictions of these two functions to $N$.

It is clear that a Poisson submanifold carries a natural Poisson bracket.

**Definition 1.10.** We call a subgroup $H$ of a Poisson—Lie group $G$ a *Poisson—Lie subgroup* if $H$ is a Poisson submanifold of $G$.

**Definition 1.11.** We call a Lie subalgebra $\mathfrak{h}$ of a Lie bialgebra $\mathfrak{G}$ a *Poisson—Lie subalgebra* if its orthogonal complement $\mathfrak{h}^\perp \subset \mathfrak{G}^*$ is an ideal with respect to the Lie algebra structure in $\mathfrak{G}^*$.

**Lemma 1.12** ([34])**.** *A connected subgroup $H$ of a Poisson—Lie group $G$ is a Poisson—Lie subgroup iff its Lie algebra $\mathfrak{h}$ is a Poisson—Lie subalgebra of $\mathfrak{G}$.*

*Remark* 1.13. We provide here some details since this lemma is a key element in many computations in this paper, and this proof carries a lot of similarities with Lemma 9.18 below. On the other hand, if a reader is not satisfied with the sketch below, we can note that this lemma is a simple corollary of Proposition 5.5, and the proof of that proposition does not depend on this lemma.

*Proof.* Call a point $h \in H$ a *good point* if the value at $h$ of a Poisson bracket of two functions on $G$ is determined by their restrictions to $H$. It is easy to see that such points form a subgroup of $H$. Moreover, the condition on $\mathfrak{h}$ implies that the bracket of a vanishing on $H$ function with any function has a zero of the second order at the unity $e \in H$. Thus the infinitely near to $e$ points are good, and we can use the fact that they generate the whole group $G$.

To utilize this argument we represent any point $h \in H$ as a product $h = h_1 h_2 \ldots h_N$ with $\operatorname{dist}(h_i, e) = O\left(\frac{1}{N}\right)$. Now the compatibility of the Poisson bracket with the group structure on $G$ implies that $\{f, g\}|_h = O\left(\frac{1}{N}\right)$ if $f|_H = 0$. $\quad\square$

*Remark* 1.14. To make this discussion less vague we should have introduced a bivector field $\eta$ corresponding to the Poisson bracket, as we do in Section 9.4. This would show that points on $\exp \mathfrak{h}$ are good, and $\exp \mathfrak{h}$ generates a dense subgroup in $H$.

**Definition 1.15.** We say that a Poisson—Lie group $G$ acts on a Poisson manifold $M$ in a *Poisson—Lie way* if the natural mapping

$$G \times M \to M$$

preserves the Poisson brackets (on the left-hand side we consider the usual direct product Poisson bracket).

*Remark* 1.16. This definition *does not imply* that this action preserves the Poisson structure on $M$. Moreover, if for some point $m \in M$ the action of $\operatorname{Lie}(G)$ on $m$, i.e., the mapping $\operatorname{Lie}(G) \to T_m M$, has no kernel, then such an action preserves the Poisson structure on $M$ *only if* the Poisson structure on $G$ is trivial. We can express the relation between the action and the Poisson structure on $G$ by a vague phrase *the action of $g \in G$ changes the Poisson structure on $M$ on amount corresponding to the* value *of the Poisson structure on $G$ at $g$.* It is a good exercise to provide an exact statement of this sense.

*Remark* 1.17. These definitions imply that a connected Lie subgroup is a Poisson—Lie subgroup iff its Lie algebra is a Poisson—Lie subalgebra. Moreover, they are compatible with the notion of Poisson—Lie action: the restriction of such an action to a Poisson—Lie subgroup is again a Poisson—Lie action. Finally, it should be mentioned that in the latter case the *quotient $M/H$* by the action of this subgroup carries a natural Poisson structure. Indeed, the Poisson bracket of two $H$-invariant functions is again $H$-invariant.[1]

In this paper we are going to apply this technique to the infinite-dimensional Lie algebra of pseudodifferential symbols (cf. [34]) and to its central extension.

## 2. The Lie algebra of pseudodifferential symbols and $\log(d/dx)$

We describe here an explicit construction of the central extension for the Lie algebra of pseudodifferential symbols on the circle ([20]). To define the corresponding 2-cocycle it turns out to be useful to introduce a concept of the logarithm of the derivative symbol.

---

[1] Actually, for this we use only the fact that $\mathfrak{h}^\perp$ is a subalgebra of $\mathfrak{G}^*$ (see proposition 8.2).

## 2.1. Algebra of symbols.

**Definition 2.1.** An associative ring $\mathfrak{g}$ of pseudodifferential symbols is the ring of formal series $A(x, D) = \sum_{-\infty}^{n} a_i(x) D^i$ with respect to $D$, where $a_i(x)$ are (real or complex, scalar or matrix) functions on the line or the circle, and the variable $D$ denotes $d/dx$. For a given symbol $A$ we call the largest number $N$ such that $a_N \neq 0$ *the order* or *the degree* $\deg A$ *of* $A$. The multiplication law in $\mathfrak{g}$ is given by the multiplication of symbols:

$$(2.1) \qquad A(x, \xi) \circ B(x, \xi) = \sum_{n \geq 0} \frac{1}{n!} A_\xi^{(n)}(x, \xi) B_x^{(n)}(x, \xi)$$

(here $A_\xi^{(n)} = \frac{d^n}{d\xi^n} A(x, \xi)$, $B_x^{(n)} = \frac{d^n}{dx^n} B(x, \xi)$), and it coincides with the usual multiplication law on the subalgebra $\mathfrak{g}_{DO} \subset \mathfrak{g}$ of *differential* operators (i.e. on polynomials with respect to $D = \xi$). This product determines the Lie algebra structure on $\mathfrak{g}$:

$$(2.2)$$
$$[A, B] = A \circ B - B \circ A = \sum_{n \geq 1} \frac{1}{n!} \left( A_\xi^{(n)}(x, \xi) B_x^{(n)}(x, \xi) - A_x^{(n)}(x, \xi) B_\xi^{(n)}(x, \xi) \right).$$

*Remark* 2.2. Formula (2.1) expresses the noncommutative multiplication of symbols $A(x, D)$ in terms of the commutative (in the scalar case) multiplication of functions $A(x, \xi)$, where we identify a symbol $\sum_{-\infty}^{n} a_i(x) D^i$ with the function $\sum_{-\infty}^{n} a_i(x) \xi^i$. We use two different letters $D$ and $\xi$ for the same variable here to avoid a confusion of commutative multiplication appearing in the right hand side of (2.1) with the noncommutative one in the left-hand side. Due to this convention we can freely drop the symbol $\circ$ anytime below: any multiplication of symbols in notations with $D$ is the $\circ$-product.

The commutative multiplication here is what physicists call the ordered product $: AB :$.

*Remark* 2.3. Here and below we can change the real coefficients to complex coefficients without any change in discussion.

There is an operator res: $\mathfrak{g} \to C^\infty(S^1)$ on the ring

$$\mathfrak{g}\colon \operatorname{res}\left(\Sigma a_i(x) D^i\right) = a_{-1}(x).$$

The main property of res is $\int \operatorname{res}[A, B] = 0$ for arbitrary $A, B \in \mathfrak{g}$ (here and below we integrate over the circle $S^1$ or over the line $\mathbb{R}^1$ in the case of rapidly decreasing coefficients). This justifies the

**Definition 2.4.** Let $\operatorname{Tr} A \overset{\mathrm{def}}{=} \int \operatorname{res} A$. Define the pairing $(\cdot, \cdot)$ on the algebra $\mathfrak{g}$ as

$$(A, B) = \operatorname{Tr}(A \circ B).$$

**Definition 2.5.** Define two Lie subalgebras $\mathfrak{g}_{DO}$ and $\mathfrak{g}_{int}$ of the Lie algebra $\mathfrak{g}$ as

$$\mathfrak{g}_{DO} = \left\{ \sum_{i=0}^{n} a_i(x) D^i, n \in \mathbb{N} \right\}, \qquad \mathfrak{g}_{int} = \left\{ \sum_{i<0} a_i(x) D^i \right\} = \{ A \in \mathfrak{g} \mid \deg A \leq -1 \}.$$

The main property of res implies that the above pairing is symmetric. The following proposition is a direct consequence of definitions:

**Proposition 2.6 (see [33]).** *The Lie algebras* $(\mathfrak{g}, \mathfrak{g}_{DO}, \mathfrak{g}_{int})$ *form a Manin triple. In other words, the above pairing is nondegenerate, and the Lie subalgebras* $\mathfrak{g}_{DO}$ *and* $\mathfrak{g}_{int}$ *are complimentary isotropic subspaces.*

*Remark* 2.7. We discuss the details of the Poisson geometry associated with this Manin triple in Section 5. Here we consider instead a remarkable central extension of $\mathfrak{g}$.

**2.2. Logarithm of the derivative operator.** Consider the formal expression $\log D$. Certainly, for any pseudodifferential symbol $A \in \mathfrak{g}$ the formal product $A \circ \log D$ (according to (2.1) and the convention that the symbol of $\log D$ is $\log \xi$) does not belong to $\mathfrak{g}$.

*Remark* 2.8. This is a big source of confusion, since one almost automatically associates the symbol $\log \xi$ with a series like $\sum_{1}^{\infty} \frac{(-1)^{k+1}}{k}(\xi - 1)^k$. However, the boundaries of summation in Definition 2.1 of an element of $\mathfrak{g}$ clearly indicate that we should consider this expansion as an expansion near $\xi = \infty$. It is clear that $\log \xi$ has no power expansion near this point.[2]

The crucial point is that the formal commutator[3]

$$[\log D, A] = \log D \circ A - A \circ \log D$$

(here the multiplication is defined by the same Formula (2.1)) is an element of $\mathfrak{g}$. Thus $\log D$ acts on $\mathfrak{g}$ by commutation $[\log D, *]$ and defines an outer derivation of $\mathfrak{g}$. In coordinate form, if $A = \sum_{i=-\infty}^{n} a_i(x) D^i$, then (due to a direct substitution into (2.2))

$$(2.3) \qquad\qquad [\log D, A] = \sum_{k \geq 1} \frac{(-1)^{k+1}}{k} A_x^{(k)} D^{-k}.$$

Note that even for a differential operator $A \in \mathfrak{g}_{DO}$ the result $[\log D, A]$ is, generally speaking, a pseudodifferential symbol.

---

[2]However, it is an adjoint element to eigenvalue $f(x, \xi) \equiv 1$ of the action of dilatations in $\xi$ (if we consider the Laurent series as a decomposition into the eigenvectors with respect to this action). Thus we can consider $\log D$ as a "symbol of close to 0 order".

[3]I.e., the result of substitution of $\log D$ into a formula (2.2) for commutator of pseudodifferential symbols.

**Theorem 2.9** ([20])**.** *The following 2-cocycle*

$$(2.4) \qquad c\left(A, B\right) = \mathrm{Tr}\left(\left[\log D, A\right] \circ B\right) = \mathrm{Tr}\left(\sum_{k \geq 1} \frac{(-1)^{k+1}}{k} A_x^{(k)} D^{-k} \circ B\right),$$

*gives a nontrivial central extension* $\widehat{\mathfrak{g}}$ *of the Lie algebra* $\mathfrak{g}$. *(Here $A$ and $B$ are arbitrary pseudodifferential symbols on $S^1$.) The restriction of this cocycle to $\mathfrak{g}_{DO}$ gives a nontrivial central extension of $\mathfrak{g}_{DO}$.*

*Remark* 2.10. The restriction of this cocycle to the subalgebra of vector fields (considered as differential operators of the first order) is the Gelfand—Fuchs cocycle of the Virasoro algebra. Indeed,

$$\begin{aligned}
c\left(f\left(x\right) D, g\left(x\right) D\right) &= \mathrm{Tr}\left(\left[\log D, f\left(x\right) D\right] \circ g\left(x\right) D\right) \\
&= \mathrm{Tr}\left(\left(f'D^0 - f''D^{-1}/2 + f'''D^{-2}/3 - \dots\right) \circ gD^1\right) \\
&= \mathrm{Tr}\left(\dots + f''\left(x\right) g'\left(x\right) D^{-1}/6 + \dots\right) \\
&= \frac{1}{6} \int f''\left(x\right) g'\left(x\right) dx.
\end{aligned}$$

This observation implies the nontriviality of the cocycle on $\mathfrak{g}$ and $\mathfrak{g}_{DO}$, since even the restriction on a subalgebra is nontrivial.

*Proof of the Theorem.* Skew symmetry of $c\left(A, B\right)$ is a consequence of the identities:

$$\left[\log D, A \circ B\right] = \left[\log D, A\right] \circ B + A \circ \left[\log D, B\right] \quad \text{and} \quad \mathrm{Tr}\left[\log D, A\right] = 0$$

for any $A, B \in \mathfrak{g}$. These identities themselves follow immediately from (2.1)-(2.3). The same identities together with the Jacobi identity on $\mathfrak{g}$ allow one to verify the cocycle property:

$$\underset{A,B,C}{\mathrm{Alt}}\, c\left(A, \left[B, C\right]\right) =$$
$$\mathrm{Tr}\left(\left[\log D, A\right] \circ \left[B, C\right] + \left[\log D, C\right] \circ \left[A, B\right] + \left[\log D, B\right] \circ \left[C, A\right]\right) = 0.$$

$\square$

*Remark* 2.11. Assume for a moment that $\log D$ were an element of the algebra $\mathfrak{g}$. Then we could define not only the commutator $\left[\log D, A\right]$ but also a product $\log D \circ A$, and rewrite the cocycle $c\left(A, B\right) = \mathrm{Tr}\left(\left[\log D, A\right] \circ B\right)$ as $c\left(A, B\right) = \mathrm{Tr}\left(\log D \circ \left[A, B\right]\right)$. The last form means that the cocycle of $c\left(A, B\right)$ is a 2-coboundary (and hence trivial cocycle) because it is a linear function of the commutator $c\left(A, B\right) = \left\langle \log D, \left[A, B\right]\right\rangle$. Recalling that $\log D \notin \mathfrak{g}$, we get a heuristic proof of the non-triviality of the cocycle.

The last expression of the cocycle $c\left(A, B\right) = \left\langle \log D, \left[A, B\right]\right\rangle$ shows that $\log D$ is in some sense an element of the dual space $\mathfrak{g}^*$. We will effectively exploit this observation further (see Section 4).

*Remark* 2.12. The value of $c\left(f\left(x\right)D^m, g\left(x\right)D^n\right)$ on the homogeneous generators of $\mathfrak{g}$ vanishes for $n + m + 1 < 0$ and, generally speaking, does not vanish for $m + n + 1 \geq 0$. The restriction of this cocycle to the Lie algebra of differential operators $\mathfrak{g}_{\mathrm{DO}}$ (i.e., $n, m \geq 0$) coincides with the Kac—Peterson cocycle [15]:

$$c\left(f\left(x\right)D^m, g\left(x\right)D^n\right) = \frac{m!n!}{(m+n+1)!} \int f^{(n)} g^{(m+1)} dx$$

(in this form it was written in [29]).

*Remark* 2.13. It should be mentioned that the subalgebra $\mathfrak{g}_{\mathrm{DO}}$, unlike $\mathfrak{g}$ itself, can be embedded into $\mathfrak{gl}_\infty$, and the cocycle on $\mathfrak{g}_{\mathrm{DO}}$ is the pullback of the only nontrivial cocycle on $\mathfrak{gl}_\infty$ [4]. However, the ambient algebra $\mathfrak{g}$ does not admit such an embedding.

*Remark* 2.14. Exactly the Lie algebra of differential operators extended by the "logarithmic" cocycle appeared during last a few years in physical papers under the name $W_{1+\infty}$, see [28, 3].

*Remark* 2.15. Notice also that the Lie algebra $\mathfrak{g}_{\mathrm{DO}}$ of differential operators on the circle has exactly one central extension [9, 25], while the Lie algebra of pseudodifferential symbols has two independent central extensions [36, 37, 10]. The formula for the second cocycle can be written in a form similar to (2.4):

$$(2.5) \qquad\qquad c^\circ\left(A, B\right) = \mathrm{Tr}\left(\left[x, A\right] \circ B\right).$$

Here $x$ is a natural coordinate on the universal covering of $S^1$, considered as a multivalued function on $S^1$. Here, as well as in the case of (2.4), $x$ is not an element of $\mathfrak{g}$, but $[x, L]$ is. Not only this cocycle generates a trivial class of cohomology of $\mathfrak{g}_{\mathrm{DO}}$, but the restriction of it to $\mathfrak{g}_{\mathrm{DO}}$ vanishes: if $A, B \in \mathfrak{g}_{\mathrm{DO}}$, then $[x, A] \circ B$ is a differential operator, therefore has zero residue.

*Remark* 2.16. Note that we can write an analogue of this cocycle $c^\circ$ in the case of pseudodifferential symbols with coefficients in Laurent polynomials. However, in this case we should change $x$ in Formula (2.5) to $\log x$. Indeed, the latter case is essentially case of functions on a small circle around $x = 0$, and the multivalued function on this circle that changes the value by 1 whenever we make a complete rotation is $\mathrm{const} \cdot \log x$.

*Remark* 2.17. In [30] the logarithmic cocycle has been generalized to the case of pseudodifferential symbols on compact manifolds. Detailed survey on various extensions of infinite-dimensional Lie algebras and groups see in [31].

### 3. The group of pseudodifferential symbols

We describe in this section the underlying Lie group and the corresponding Lie algebra of the main Poisson—Lie group we will work with: the Lie group of pseudodifferential symbols of real (or complex) order. The elements of this group are usually called "symbols of classical pseudodifferential symbols".

### 3.1. Classical symbols.

**Definition 3.1.** A classical pseudodifferential symbol $(\Psi DS)$ is an expression of the form $L = \left( \sum_{k=-\infty}^{0} u_k(x) D^k \right) \circ D^t$, where $t \in \mathbb{R}$, $D = d/dx$, and $u_k(x)$ are functions on $\mathbb{R}$ or $S^1$. The multiplication of the symbols is uniquely defined by the commutation relation

$$\left[ D^t, u(x) \right] = \sum_{l \geq 1} \binom{t}{l} u^{(l)}(x) D^{t-l},$$

where $\binom{t}{l} = \frac{t(t-1)\ldots(t-l+1)}{l!}$. Call the (real) number $t$ the *degree* or the *order* of the symbol $L$.

*Remark* 3.2. We could as well define the product of two symbols by the same Formula (2.1) as above. Moreover, if we consider symbols with complex coefficients, we can allow complex values for the parameter $t$ as well. Below we will not explicitly specify whether we work with real or complex coefficients and degrees, since in the questions we discuss here there is no difference between these two cases.

Define on the set of $\Psi DS$ the topology as the usual $C^\infty$-topology in the direction of the variable on $S^1$ (or $\mathbb{R}$) and the topology of projective limit along the variable $k$. For an individual $k$ we consider a usual $C^\infty(S^1)$ (or $C^\infty(\mathbb{R})$) topology on the coefficient $u_k(x)$. Then the basic neighborhoods of a point $L^{(0)}$ are the sets of $L$'s such that $|t - t_0| < \varepsilon$, $|u^{(k)}(x) - u_0^{(k)}(x)| < \phi(x)$, $k = 0, \ldots, l$ for fixed $\varepsilon, l$ and a fixed positive function $\phi(x)$.

Notice that two different symbols (distinguished by integer parts of $t$) may correspond to the same object. However, in the following definition the notion of the degree is well defined.

**Definition 3.3.** The group $\widetilde{G}_{\text{int}}$ consists of the $\Psi DS$'s with the leading coefficient $u_0(x) \equiv 1$.

It is simple to check that this is a group indeed. Our purpose now is to apply the Poisson—Lie formalism to the infinite-dimensional group $\widetilde{G}_{\text{int}}$. We pretend that any result valid for finite-dimensional Poisson—Lie group has corresponding counterpart for the group in question.

We begin with determining the Lie algebra corresponding to the Lie group $\widetilde{G}_{\text{int}}$. For the subgroup $G_{\text{int}}$ of symbols of degree 0 (i.e., for $L = 1 + \sum_{k=-\infty}^{-1} u_k(x) D^k$) this is a straightforward task: the Lie algebra $\mathfrak{g}_{\mathrm{int}}$ consists of all symbols $P = \sum_{k=-\infty}^{-1} a_k(x) D^k$ of degree $-1$ (with an arbitrary leading coefficient $a_{-1}$).

We call a symbol from $\mathfrak{g}_{\mathrm{int}}$ a *symbol of an integral operator*, or an *integral symbol*.

### 3.2. Exponential map for integral symbols.
Let us consider the structure of the exponential map in this case. Since later we consider the more complicated case of the group $\widetilde{G}_{\mathrm{int}}$, let us investigate the group $G_{\mathrm{int}}$ for a while.

**Proposition 3.4.** *The exponential mapping $\mathfrak{g}_{int} \to G_{int}$ is well-defined and surjective for both the periodical case and the case of coefficients on a line.*

*Proof.* Fix a pseudodifferential symbol $P$ of degree $-1$. The would-be exponent $L_{(s)}$ of $s\,P$ should satisfy the equation

$$(3.1) \qquad \left(\frac{d}{ds}L_{(s)}\right) \circ \left(L_{(s)}\right)^{-1} = P,$$

or

$$\frac{d}{ds}L_{(s)} = P \circ L_{(s)}.$$

Rewriting this in terms of coefficients of the symbol

$$L_{(s)} = 1 + \sum_{k=-\infty}^{-1} u_{k(s)}(x) D^k$$

we get a "triangle system" of the form

$$(3.2) \qquad \frac{d}{ds}u_{k(s)}(x) = \Phi_k\left(u_{1(s)}, \ldots, u_{k-1(s)}\right)(x), \quad k \geq 1$$

where $\Phi_k$ is a differential polynomial in $u_1, \ldots, u_{k-1}$, (recall that it is a polynomial in $u_1, \ldots, u_{k-1}$ and their derivatives:

$$(3.3) \qquad \Phi_k\left(u_1, \ldots, u_{k-1}\right) = \sum_M \Phi_{k,M} \prod_{i,j} \left(u_i^{(j)}\right)^{m_{ij}},$$

here $M$ denotes a multiindex).

Equation (3.2) can be solved in quadratures (for initial condition $u_k|_{s=0} = 0$ for all $k < 0$, for instance), since the right-hand side of (3.2) does not depend on $u_k$. The solution is periodic in $x$ (respectively rapidly decreasing) if the initial condition and $P$ were those. Hence the exponential map is correctly defined. $\square$

*Remark* 3.5. This argument can be made explicit by considering the exponential series $\exp(s\,P) = \sum_{k \geq 0} \frac{1}{k!} s^k P^k$. This series converges because $\deg P^k = -k$, therefore to compute a certain coefficient of $\exp s\,P$ we have to calculate only a finite number of summands. Now, in order to show that the exponential map is surjective we need to consider the series for the logarithm: $\log{(1+P)} = \sum_{k \geq 1} \frac{(-1)^k}{k} P^k$ that converges by the same reason.

**3.3.** $\frac{d}{ds}|_{s=0} D^s = \log D$.   Now we go back to the group $\widetilde{G}_{int}$ of pseudodifferential symbols of an arbitrary real order. It contains the considered above group $G_{int}$ as a subgroup of codimension one.

**Proposition 3.6.** *The Lie algebra* $\widetilde{\mathfrak{g}}_{int}$ *for the Lie group* $\widetilde{G}_{int}$ *is an extension of the Lie algebra of integral symbols* $\mathfrak{g}_{int}$ *by the formal symbol* $\log D$. *That means that*

(1) $\widetilde{\mathfrak{g}}_{int} = \mathfrak{g}_{int} \oplus \{\lambda \log D\}$ *as a linear space;*
(2) *the commutator of two integral symbols is standard, see* (2.2);
(3) $[\log D, \log D] = 0$; *and*
(4) *the commutation relation for* $\log D$ *and* $P = u_j(x) D^j \in \mathfrak{g}_{int}$ *is given by Formula* (2.3):

$$[\log D, P] = \sum_{l=1}^{\infty} \frac{(-1)^{l+1}}{l} u_j^{(l)} D^{j-l}.$$

*Remark* 3.7. In fact we defined just the extension $\langle \log D \rangle \ltimes \mathfrak{g}_{int}$ of the Lie algebra $\mathfrak{g}_{int}$ by the outer derivative $\log D$. The definition of commutator in this algebra is just a restatement of Formulae (2.1)–(2.2).

*Proof.* The group $G_{int}$ is a normal subgroup of $\widetilde{G}_{int}$ (and thus the corresponding Lie algebra $\mathfrak{g}_{int}$ is a subalgebra in $\widetilde{\mathfrak{g}}_{int}$). Let us consider the complementary 1-parameter subgroup $\{D^s\}$ (the group $\widetilde{G}_{int}$ is a semidirect product $\{D^s\} \ltimes G_{int}$ of these two subgroups). Computing the commutation relations of the tangent elements to the subgroups $D^s$ and $G_{int}$ we get the following:

$$D^{\varepsilon} \cdot \exp{(\delta P)} \cdot D^{-\varepsilon} \cdot \exp{(-\delta P)} \approx D^{\varepsilon} (1 + \delta P) D^{-\varepsilon} (1 - \delta P) \approx 1 + \delta \left( D^{\varepsilon} P D^{-\varepsilon} - P \right)$$

$$= 1 + \delta \left[ D^{\varepsilon}, P \right] D^{-\varepsilon} \approx 1 + \delta \left[ D^{\varepsilon}, u_j D^j \right] D^{-\varepsilon}$$

$$= 1 + \delta \sum_{l=1}^{\infty} \binom{\varepsilon}{l} u_j^{(l)} D^{j-l}$$

$$\approx 1 + \delta \varepsilon \frac{\partial}{\partial \varepsilon}|_{\varepsilon=0} \sum_{l=1}^{\infty} \binom{\varepsilon}{l} u_j^{(l)} D^{j-l},$$

here $P = u_j(x) D^j$, $\binom{\varepsilon}{l} = \frac{\varepsilon(\varepsilon-1)...(\varepsilon-l+1)}{1 \cdot 2 \cdot ... \cdot l}$, and $\approx$ denotes an equality modulo terms $O\left(\varepsilon^2, \delta^2\right)$ for $\varepsilon, \delta \to 0$.

Since $\frac{\partial}{\partial \varepsilon}|_{\varepsilon=0} \binom{\varepsilon}{l} = \frac{(-1)^{l+1}}{l}$ we come to

$$D^{\varepsilon} \cdot \exp{(\delta P)} \cdot D^{-\varepsilon} \cdot \exp{(-\delta P)} \approx 1 + \delta \varepsilon \sum_{l=1}^{\infty} \frac{(-1)^{l+1}}{l} u_j^{(l)} D^{j-l} = 1 + \delta \varepsilon \left[ \log D, P \right].$$

Thus the formal symbol $\log D$ is a tangent vector to the subgroup $D^s$, and the Lie algebra of $\widetilde{G}_{\mathrm{int}}$ can be naturally identified with the algebra $\widetilde{\mathfrak{g}}_{\mathrm{int}}$.  $\square$

*Remark* 3.8. Heuristically, the tangent vector to $D^s$ can be obtained by differentiation of this 1-parameter subgroup with respect to $s$ at $s = 0$:

$$\frac{d}{ds}\big|_{s=0} D^s = \log D \cdot D^s |_{s=0} = \log D.$$

**3.4. Exponential map.**    Now we can describe the structure of the exponential map $\exp : \widetilde{\mathfrak{g}}_{\mathrm{int}} \to \widetilde{G}_{\mathrm{int}}$. We already know it for elements $P \in \mathfrak{g}_{\mathrm{int}}$ (with vanishing coefficient at $\log D$), so for complete consideration it is sufficient to compute $\exp\left(s\left(\log D + P\right)\right) = L_{(s)}$.

The description by exponential series does not work now (since it does not work for $\exp\left(t \log D\right) = D^t$), so to define "exp" we again have to consider the equation

$$(3.4) \qquad\qquad \frac{dL_{(s)}}{ds}\left(L_{(s)}\right)^{-1} = \log D + P$$

**Proposition 3.9.** *The exponential map* $\exp : \widetilde{\mathfrak{g}}_{int} \to \widetilde{G}_{int}$ *given by the solution of* (3.4)

$$s\left(\log D + P\right) \mapsto L_{(s)}$$

*is well defined on the entire Lie algebra* $\widetilde{\mathfrak{g}}_{int}$ *and is a surjective mapping onto* $\widetilde{G}_{int}$.

*Proof.* Let $L_{(s)} = \left(1 + P_{(s)}\right) D^s = \left(1 + \sum_{j=-\infty}^{-1} u_{j(s)} D^j\right) D^s$. First of all we show that Equation (3.4) is equivalent to the following one:

$$\frac{d}{ds} P_{(s)} = \left[\log D, P_{(s)}\right] + P \circ \left(1 + P_{(s)}\right).$$

Indeed, replacing the differentiation of $L_{(s)}$ by a finite difference we get

$$\varepsilon \frac{dL_{(s)}}{ds}\left(L_{(s)}\right)^{-1} + O\left(\varepsilon^2\right) = L_{(s+\varepsilon)}\left(L_{(s)}\right)^{-1} = \left(1 + P_{(s+\varepsilon)}\right) D^{s+\varepsilon} D^{-s}\left(1 + P_{(s)}\right)^{-1}$$

$$= \left(1 + P_{(s+\varepsilon)}\right) D^\varepsilon \left(1 + P_{(s)}\right)^{-1}$$

$$= \left(D^\varepsilon \left(1 + P_{(s+\varepsilon)}\right) - \left[D^\varepsilon, P_{(s+\varepsilon)}\right]\right)\left(1 + P_{(s)}\right)^{-1}$$

$$= \left(\left(D^\varepsilon - 1\right)\left(1 + P_{(s+\varepsilon)}\right) + \left(1 + P_{(s+\varepsilon)}\right) - \left[D^\varepsilon, P_{(s+\varepsilon)}\right]\right)\left(1 + P_{(s)}\right)^{-1}$$

$$= \varepsilon \log D + \varepsilon \frac{dP_{(s)}}{ds}\left(1 + P_{(s)}\right)^{-1} - \varepsilon\left[\log D, P_{(s)}\right]\left(1 + P_{(s)}\right)^{-1} + O\left(\varepsilon^2\right)$$

Thus (3.4) is equivalent to the equation

$$P = \frac{dP_{(s)}}{ds}\left(1 + P_{(s)}\right)^{-1} - \left[\log D, P_{(s)}\right]\left(1 + P_{(s)}\right)^{-1}$$

or

$$(3.5) \qquad \frac{d}{ds}P_{(s)} = \left[\log D, P_{(s)}\right] + P \circ \left(1 + P_{(s)}\right).$$

The last equation has the same form (3.2) as (3.1), and therefore it can be solved in quadratures. Hence the exponential map is defined correctly on $\widetilde{\mathfrak{g}}_{\text{int}}$.

Let us prove a surjectivity of the exponential map. To invert "exp" we need to determine the element "log"$\left((1 + P) \circ D^t\right)$, where $(1 + P) \circ D^t \in \widetilde{G}_{\text{int}}$ (here $P = \Sigma_{j=-\infty}^{-1} u_j D^j \in \mathfrak{g}_{\text{int}}$). Surjectivity of the exponential map $\mathfrak{g}_{\text{int}} \to G_{\text{int}}$ for the non-extended subalgebra of integral symbols (Proposition 3.4) allows us to find $\bar{P} \in \mathfrak{g}_{\text{int}}$ such that $1 + P = \exp\left(\bar{P}\right)$ and thus

$$(1 + P) D^t = \left(\exp \bar{P}\right) \circ \exp\left(t \log D\right).$$

Now using the Campbell-Hausdorff formula

$$\exp A \circ \exp B = \exp\left(A + B + \frac{[A, B]}{2} + \frac{1}{12}\left([A\,[A, B]] + [B\,[B, A]]\right) + \dots\right)$$

we get

$$\begin{aligned}
\text{"log"}\left((1 + P) \circ D^t\right) &= \text{"log"}\left(\exp \bar{P} \circ \exp\left(t \log D\right)\right) \\
&= t \log D + \bar{P} + t\left[\bar{P}, \log D\right] + \dots
\end{aligned}$$

Notice that the terms in the final expression have decreasing order (going to $-\infty$) due to increasing number of commutators. Hence any coefficient at $D^j$ for "log"$\left(P \circ D^t\right)$ is defined by a finite number of the terms and thus the inverse of the exponential map is well defined.   ☐

*Remark* 3.10. Propositions 3.4 and 3.9 show that the (extended) Lie group $\widetilde{G}_{\text{int}}$ and the (extended) Lie algebra $\widetilde{\mathfrak{g}}_{\text{int}}$ of pseudodifferential operators are very "suitable" objects to deal with. The group $\widetilde{G}_{\text{int}}$ is an infinite-dimensional analogue of a unipotent group. Analogously to the finite-dimensional case where the exponential map is one-to-one and the exponential series consists of a finite number of terms, in our situation every coefficient in $L = \exp\left(\log D + P\right)$ is defined by a finite number of terms of the integral symbol $P$. It is not very surprising, since the algebra in question is quasinilpotent.

We would like to emphasize that the Lie algebra of differential operators does not have even a shadow of these properties: no hope for existence of the corresponding group, and therefore for any kind of exponential map. Even for the subalgebra $\text{Vect} = \left\{u_1\left(x\right)D\right\}$ of vector fields its image under the exponential map in the group $\text{Diff}\left(S^1\right)$ of all diffeomorphisms of the circle generates a nowhere dense normal subgroup of $\text{Diff}\left(S^1\right)$.

*Remark* 3.11. The crucial observation for the applications of this group to $n$-KdV hierarchies lies in the fact that this group contains a semigroup of differential operators.

*Remark* 3.12. Proposition 3.6 shows how to extend the Lie algebra $\mathfrak{g}_{\mathrm{int}}$ by an element $\log D$. However, it is easy to see that the commutator with $\log D$ is a derivative of $\mathfrak{g}$ as well, therefore the same formulae allow one to define the extension $\mathfrak{g}_{\log}$ of the Lie algebra $\mathfrak{g}$ by an element $\log D$. The remarkable property of this extension is its *duality* to the central extension by the cocycle $c$ discussed in Section 2. This duality is with respect to the natural pairing $(\cdot, \cdot)$ on $\mathfrak{g}$. In the following section we will see that it is possible to strengthen this (vague) duality to a definition of a much more interesting object.

*Remark* 3.13. The good properties of the exponential mapping make it possible to define the mapping $\log \colon \widetilde{G}_{\mathrm{int}} \to \widetilde{\mathfrak{g}}_{\mathrm{int}}$. Obviously, this notion of $\log L$, $L \in \widetilde{G}_{\mathrm{int}}$ is compatible with the (symbolic) notation $\log D$ for the additional element in $\widetilde{\mathfrak{g}}_{\mathrm{int}}$.

## 4. The extended algebra of pseudodifferential symbols as a Lie bialgebra

**4.1. Double extension.**   Here we do the last step in definitions of the Poisson—Lie structure on the Lie group from Section 3: we define an ambient Lie algebra for a Manin triple for $\widetilde{\mathfrak{g}}_{\mathrm{int}}$. Recall that in Section 2 we have shown that nonextended algebra $\mathfrak{g}_{\mathrm{int}}$ can be included into the Manin triple $(\mathfrak{g}, \mathfrak{g}_{\mathrm{DO}}, \mathfrak{g}_{\mathrm{int}})$. Here $\mathfrak{g}$ is the Lie algebra of pseudodifferential symbols equipped with a nondegenerate ad-invariant bilinear form $(A, B) = \mathrm{Tr}\, AB$, and the Lie algebra $\mathfrak{g}_{\mathrm{DO}}$ of differential operators is dual to $\mathfrak{g}_{\mathrm{int}}$ with respect to this pairing.

As a vector space the ambient Lie algebra of the would-be Manin triple for $\widetilde{\mathfrak{g}}_{\mathrm{int}}$ is isomorphic to $\widetilde{\mathfrak{g}}_{\mathrm{int}} \oplus \widetilde{\mathfrak{g}}_{\mathrm{int}}^*$. Since $\widetilde{\mathfrak{g}}_{\mathrm{int}}$ is "one dimension bigger" than $\mathfrak{g}_{\mathrm{int}}$, this ambient Lie algebra should be two dimensions bigger than the ambient Lie algebra of the Manin triple $(\mathfrak{g}, \mathfrak{g}_{\mathrm{DO}}, \mathfrak{g}_{\mathrm{int}})$ for $\mathfrak{g}_{\mathrm{int}}$. We know already two different Lie algebras that are both one dimension bigger than $\mathfrak{g}$. The first is the central extension $\widehat{\mathfrak{g}}$ by the cocycle $c$ from Section 2. The second is the extension $\mathfrak{g}_{\log}$ by $\log D$ from Section 3. We have already noted that these two extensions are dual to each other. This means that if one could carry out the additions of both these elements to $\mathfrak{g}$ the resulting object would have a good chance to be self-dual (as the ambient Lie algebra of a Manin triple should be).

**Lemma 4.1 (cf. [14]).** *Let $\mathfrak{G}$ be a Lie algebra over $\mathbb{K}$ with an ad-invariant bilinear pairing $(,)$. Let $d$ be a derivation of $\mathfrak{G}$ that is skew-symmetric with respect to $(,)$. Let $\mathbb{K}C$ be the dual space to $\mathbb{K}d$. Then the vector space*

$$\mathfrak{G}_{de} = \mathfrak{G} \oplus \mathbb{K}d \oplus \mathbb{K}C$$

*carries a natural structure of a Lie algebra with ad-invariant bilinear form* $(,)_{de}$. *Here the only non-zero commutators in* $\mathfrak{G}_{de}$ *are*

$$[X_1, X_2]_{de} = [X_1, X_2] + (d(X_1), X_2) C,$$
$$[d, X] = d(X)$$

$$X, X_1, X_2 \in \mathfrak{G},$$

*the inner product remains the same on* $\mathfrak{G}$, *the subspace* $\mathfrak{G}$ *is orthogonal to* $\mathbb{K}d \oplus \mathbb{K}c$, *and* $(d, d)_{de} = (C, C)_{de} = 0$, $(d, C)_{de} = 1$.

We call $\mathfrak{G}_{de}$ *a double extension* of the Lie algebra $\mathfrak{G}$. Let $\mathfrak{g}$ be the Lie algebra of pseudodifferential symbols.

**Definition 4.2.** Let $\widetilde{\mathfrak{g}}$ to be the "double extension" of $\mathfrak{g}$, i.e. the algebra $\mathfrak{g}$ extended by the 2-cocycle (2.4) and by the symbol $\log D$:

$$\widetilde{\mathfrak{g}} = \left\{ \left( \sum_{j=-\infty}^{n} a_j(x) D^j + \lambda \log D, \gamma \right) \right\}.$$

Here $a_j$ are periodic or rapidly decreasing functions, $\lambda$ and $\gamma$ are numbers.

Two defined above extensions of the Lie algebra $\mathfrak{g}$: the extension $\mathfrak{g}_{\log D}$ and the central extension $\widehat{\mathfrak{g}}$ given by (2.4), are obviously subalgebras of this algebra.

*Remark* 4.3. Recall that in the theory of affine Lie algebras there is an absolutely analogous construction, in which one centrally extends the algebra of matrix functions on $S^1$ and adds some element $t$ to the resulting Lie algebra.[4] Such a double extension (as well as the initial algebra of matrix functions) carries a natural nondegenerate pairing.

*Remark* 4.4. To avoid confusion we recall two used above notions of extensions. A (right) *extension* of a Lie algebra $\mathfrak{G}$ by a Lie algebra $\mathfrak{H}$ is a Lie algebra $\bar{\mathfrak{G}}$ with an exact sequence

$$0 \to \mathfrak{G} \to \bar{\mathfrak{G}} \to \mathfrak{H} \to 0$$

and a Lie algebra morphism of *splitting* $\mathfrak{H} \hookrightarrow \bar{\mathfrak{G}}$ of the projection $\bar{\mathfrak{G}} \to \mathfrak{H}$. We can write an element of $\bar{\mathfrak{G}}$ as a sum of two elements: one from $\mathfrak{G}$ and the other from the image of the splitting. In this case there is a mapping $\mathfrak{H} \to \mathrm{Out}(\mathfrak{G})$ into the Lie algebra of outer derivations, and this mapping determines $\bar{\mathfrak{G}} = \mathfrak{G} \rtimes \mathfrak{H}$ up to isomorphism.

On the other hand, a *central extension* of $\mathfrak{G}$ by an abelian Lie algebra $\mathfrak{A}$ is a Lie algebra $\widehat{\mathfrak{G}}$ with an exact sequence

$$0 \to \mathfrak{A} \to \widehat{\mathfrak{G}} \to \mathfrak{G} \to 0$$

---

[4]The inner product on the currents algebra associates this element $t$ with the operator $\frac{d}{dx}$. However, the inner product in our situation will associate it with $\log \frac{d}{dx}$.

such that $\mathfrak{A} \subset \mathrm{Cent}\,\widehat{\mathfrak{G}}$. Such an extension is determined by a mapping $H_2\,(\mathfrak{G}) \to \mathfrak{A}$. We write an element of $\widehat{\mathfrak{G}}$ as $(X, \gamma)$, where $X$ is from $\mathfrak{G}$ (strictly speaking, from an image of some splitting $\mathfrak{G} \to \widehat{\mathfrak{G}}$), and $\gamma$ is the element of $\mathfrak{A}$ (in the case of $\dim \mathfrak{A} = 1$, $\mathfrak{A} = \langle C \rangle$, $\gamma$ is the coefficient at $C$).

Now the mentioned above duality is the fact that two exact sequences

$$0 \to \mathfrak{g} \to \mathfrak{g}_{\log} \to \langle \log D \rangle \to 0$$

and

$$0 \to \langle C \rangle \to \widehat{\mathfrak{g}} \to \mathfrak{g} \to 0$$

are dual to each other with respect to the natural pairing on $\mathfrak{g}$ (we mean that one can extend the pairing on $\mathfrak{g}$ to a pairing between $\mathfrak{g}_{\log}$ and $\widehat{\mathfrak{g}}$ that makes these two sequences dual).

Here we have collected these two diagrams together and, using self-duality of $\mathfrak{g}$, have constructed a self-dual extension $\widetilde{\mathfrak{g}}$ of $\mathfrak{g}$ that is simultaneously a central extension of the right extension $\mathfrak{g}_{\log}$ and a right extension of the central extension $\widehat{\mathfrak{g}}$. This is a reason to call such an object a *double extension*.

**4.2. Manin triple.** The algebra $\mathfrak{g}$ as a linear space is a direct sum of two natural subalgebras: $\mathfrak{g}_+ = \mathfrak{g}_{\mathrm{DO}}$ consisting of differential operators $\sum_{j \geq 0} a_j D^j$, and $\mathfrak{g}_- = \mathfrak{g}_{\mathrm{int}}$ consisting of integral symbols $\sum_{j=-\infty}^{-1} a_j D^j$.

The crucial observation is that the algebra $\widetilde{\mathfrak{g}}$ also has two remarkable subalgebras:

(1) $\widetilde{\mathfrak{g}}_+ = \widehat{\mathfrak{g}}_{\mathrm{DO}}$, which is the central extension of the algebra of differential operators: $\left( \sum_{j \geq 0} a_j D^j, \gamma \right)$, where $\gamma$ is dual to the 2-cocycle

$$c\,(M, N) = \mathrm{Tr}\,([\log D, M] \circ N)\,,$$

and

(2) $\widetilde{\mathfrak{g}}_-$, which is the algebra $\widetilde{\mathfrak{g}}_{\mathrm{int}}$ consisting of integral symbols together with $\log D$: $\sum_{j=-1}^{-\infty} a_j D^j + \lambda \log D$.

**Proposition 4.5.** $(\widetilde{\mathfrak{g}}, \widetilde{\mathfrak{g}}_+, \widetilde{\mathfrak{g}}_-)$ *is a Manin triple (or, equivalently, $\widetilde{\mathfrak{g}}_{int} = \widetilde{\mathfrak{g}}_-$ is a Lie bialgebra).*

*Proof.* As Lemma 4.1 shows, the algebra $\widetilde{\mathfrak{g}}$ (as well as $\mathfrak{g}$) has ad-invariant nondegenerate inner product ("Killing form"):

$$((A + \lambda \log D, \gamma), (B + \mu \log D, \delta)) = \mathrm{Tr}\,(A \circ B) + (\lambda \delta + \mu \gamma)$$

for $A, B \in \mathfrak{g}$. On the other hand, as a linear space $\widetilde{\mathfrak{g}} = \widetilde{\mathfrak{g}}_+ \oplus \widetilde{\mathfrak{g}}_-$, and both subalgebras are isotropic with respect to this Killing form. $\quad\square$

*Remark* 4.6. In order to verify directly that the commutator $[x_1 + \alpha_1, x_2 + \alpha_2]_{\widetilde{\mathfrak{G}}}$ of Remark 1.8 coincides with the ordinary commutation of pseudodifferential symbols, notice that the terms $\mathrm{ad}^*_{\alpha_i} x_j$ are the projections of commutators of $\Psi DS$'s to the differential operators, while $\mathrm{ad}^*_{x_i} \alpha_j$ are the projections to the integral symbols.

**Corollary 4.7.** *The Lie group $\widetilde{G}_{int}$ corresponding to the Lie bialgebra $\widetilde{\mathfrak{g}}_{int} = \widetilde{\mathfrak{g}}_-$ has a natural Lie—Poisson structure.*

This would follow immediately from Proposition 4.5 and Theorem 1.4 [26] if we could define a Lie group corresponding to the whole Lie algebra $\widetilde{\mathfrak{g}}$. Fortunately, the formulae for the Poisson—Lie group structure use only the group structure of $\widetilde{G}_-$ and the Lie-algebraic structure of $\widetilde{\mathfrak{g}}_+$ (see [26]). So we can apply general techniques of the papers [26] and [34] to obtain various advanced consequences of Proposition 4.5 for the Gelfand—Dickey structures on pseudodifferential symbols. We do this in Section 6.

*Remark* 4.8. Note that the summand $\mathfrak{g}_+$ (as well as $\widetilde{\mathfrak{g}}_+$) does not have the corresponding Lie group. (It is possible to show that its adjoint and coadjoint orbits are "fairly bad": they are not smooth or even separable while the existence of a Lie group would imply these properties [18]).

*Remark* 4.9. While the summand $\widehat{\mathfrak{g}}_{DO}$ of the decomposition $\widetilde{\mathfrak{g}} = \widehat{\mathfrak{g}}_{DO} \oplus \widetilde{\mathfrak{g}}_{int}$ is invariant with respect to commutation with $\mathrm{Vect}\,(S^1)$, the summand $\widetilde{\mathfrak{g}}_{int}$ is not.[5] Therefore, if we consider $S^1$ as an abstract 1-dimensional oriented manifold,

(1) The Lie algebra $\widehat{\mathfrak{g}}_{DO}$ is canonically defined;
(2) The Lie bialgebra structure on it is not invariant with respect to the group $\mathrm{Diff}\,S^1$ of diffeomorphisms of $S^1$;
(3) The Lie algebra $\widetilde{\mathfrak{g}}_{int}$ and the Lie group $\widetilde{G}_{int}$ are defined only after a choice of coordinate system or locally affine structure on $S^1$;
(4) After a choice of a coordinate on a circle the Lie bialgebra structure and the Poisson—Lie structures on these spaces are canonically defined.

We can finish this section by the note that the Lie subalgebra $\mathcal{D}^{(1)}$ of differential operators of order $\leq 1$ is Poisson—Lie subalgebra of $\mathfrak{g}_{DO}$. The same is true for the central extensions $\widehat{\mathcal{D}}^{(1)} \subset \widehat{\mathfrak{g}}_{DO}$. Therefore this subalgebra acts in a Poisson—Lie way both on the Poisson—Lie algebra $\widehat{\mathfrak{g}}_{DO}$ and on Poisson—Lie group $\widetilde{G}_{int}$ by adjoint and dressing action (see Section 7.3). However, this action is very different from the action of $\mathrm{Vect}\,(S^1)$. Since the Poisson—Lie structure on $\mathcal{D}^{(1)}$ is non-trivial, this action does not preserve the Poisson structures, and the (dressing) action on $\widetilde{G}_{int}$ is not the adjoint action. Moreover, the subalgebra $\widehat{\mathrm{Vect}\,(S^1)} \subset \widehat{\mathcal{D}}^{(1)}$ is not a Poisson—Lie subalgebra.

---

[5] In the case $\mathfrak{g} = \mathfrak{g}_{DO} \oplus \mathfrak{g}_{int}$ both the summands were invariant relative to the action of $\mathrm{Diff}\,(S^1)$.

## 5. The Poisson—Lie group structure on pseudodifferential symbols

In this section we show how the general Poisson—Lie group techniques (see [6, 34], and [26]) can be applied to the group of pseudodifferential symbols. In particular, as a corollary of these constructions we obtain the Gelfand—Dickey structures on the symbols and investigate their Poisson properties.

**5.1. From Manin triple to a Poisson—Lie structure.** First of all, following [26] and [34] we recall the construction of the Poisson structure on the Lie group corresponding to a Lie bialgebra.

Let $(\mathfrak{G}, \mathfrak{G}_-, \mathfrak{G}_+)$ be a Manin triple and $(\cdot, \cdot)$ be an invariant inner product on $\mathfrak{G}$. Define a bivector $r \in \mathfrak{G} \wedge \mathfrak{G}$ by the formula

$$(5.1) \qquad \langle r, \widetilde{a}_+ \wedge \widetilde{a}_- \rangle = - \langle r, \widetilde{a}_- \wedge \widetilde{a}_+ \rangle = (a_+, a_-) \,.$$

Here $a_\pm \in \mathfrak{G}_\pm$; by $\widetilde{a}$ we denote the element of $\mathfrak{G}^*$ dual to $a \in \mathfrak{G} : \langle \widetilde{a}, \cdot \rangle = (a, \cdot)$, $\langle \cdot, \cdot \rangle$ being the natural pairing between dual spaces. If we identify the space $\mathfrak{G} \otimes \mathfrak{G}$ with $\mathrm{Hom}\,(\mathfrak{G}, \mathfrak{G})$ using the inner product on $\mathfrak{G}$, then $r$ corresponds to the skewsymmetric operator $\widetilde{r}$ such that $\widetilde{r}|_{\mathfrak{G}_-} = -1$, $\widetilde{r}|_{\mathfrak{G}_+} = 1$.

Let $G$, $G_+$, and $G_-$ be the Lie groups corresponding to the Lie algebras $\mathfrak{G}$, $\mathfrak{G}_+$, and $\mathfrak{G}_-$, let also $\mathcal{R}_g$ and $\mathcal{L}_g$ be the operations of right and left multiplication by $g \in G$. Denote the corresponding mappings of the tangent spaces and their squares by

$$\mathcal{R}_{g*}, \mathcal{L}_{g*} \colon T_e \mathfrak{G} \to T_g \mathfrak{G}, \qquad \Lambda^2 \mathcal{R}_{g*}, \Lambda^2 \mathcal{L}_{g*} \colon \Lambda^2 T_e G \to \Lambda^2 T_g G.$$

Consider $r$ as an element of $\Lambda^2 \left( T_e G \right) = \Lambda^2 \mathfrak{G}$.

**Proposition 5.1** ([6, 26, 34]). *The bivector field $\eta$ on $G$ given by*

$$\eta|_g = \left( \Lambda^2 \mathcal{R}_{g*} - \Lambda^2 \mathcal{L}_{g*} \right) r$$

*defines a Poisson—Lie structure on $G$.*

**Corollary 5.2.** *The subgroup $G_-$ is a Poisson submanifold.*

*Proof.* This is an immediate corollary of Lemma 1.12. $\quad\square$

*Remark* 5.3. In the case of a Poisson structure associated with an $r$-matrix (5.1) this corollary can be verified directly. (We will do it to avoid a dead loop in the proof of Lemma 1.12.)

Indeed, let $g \in G_-$, $\alpha, \beta \in T_g^* G$, such that $\alpha \in (T_g G_-)^\perp \subset T_g^* G$. Denote $\widetilde{\alpha} = \mathcal{R}_g^* \alpha$, $\widetilde{\beta} = \mathcal{R}_g^* \beta$, $\widetilde{\alpha}, \widetilde{\beta} \in (T_e G_-)^\perp \subset \mathfrak{G}^*$. Now by definition

$$\langle \eta|_g, \alpha \wedge \beta \rangle = \left\langle r\left(\widetilde{\alpha}\right), \widetilde{\beta} \right\rangle - \left\langle r\left(\mathrm{Ad}_g^* \widetilde{\alpha}\right), \mathrm{Ad}_g^* \widetilde{\beta} \right\rangle \,.$$

Here we consider $r \in \mathfrak{G} \otimes \mathfrak{G}$ as a mapping $\mathfrak{G}^* \to \mathfrak{G}$. However, since $\widetilde{\alpha} \in \mathfrak{G}_-^\perp \subset \mathfrak{G}^*$, and $\mathrm{Ad}_g$ preserves $\mathfrak{G}_-$, $\mathrm{Ad}_g^* \widetilde{\alpha} \subset \mathfrak{G}_-^\perp$. Since the natural pairing on $\mathfrak{G}$ is Ad-invariant, we can identify $\widetilde{\alpha}$ and $\operatorname{Ad}_g^* \widetilde{\alpha}$ with elements of $\mathfrak{G}_-$. Hence both $r(\widetilde{\alpha}) = -\widetilde{\alpha}$ and $r\left(\operatorname{Ad}_g^* \widetilde{\alpha}\right) = -\operatorname{Ad}_g^* \widetilde{\alpha}$. Therefore $\langle \eta|_g, \alpha \wedge \beta \rangle = 0$, since the operator $\operatorname{Ad}_g$ is orthogonal with respect to the bracket (,). Thus the Poisson bracket $\{\varphi_1, \varphi_2\}$ of any two functions is well-defined by the restriction $\varphi_1|_{G_-}$ and $\varphi_2|_{G_-}$.

*Remark* 5.4. In Section 1 we defined the Lie bialgebra structures on both subalgebras $\mathfrak{G}_\pm$. This determines the Poisson—Lie group structure for both corresponding groups $G_\pm$. Here we have defined a Poisson—Lie structure on the ambient group $G$ and now consider those Poisson—Lie structures on $G_\pm$ as structures of Poisson—Lie subgroups of $G$.

Let $\alpha$ and $\beta$ be cotangent vectors to $G_-$ at a point $g \in G_- \left(\alpha, \beta \in T_g^* G_-\right)$. Extend them arbitrarily to cotangent vectors (at $g$) to the large group $G \supset G_-$ (we denote these extensions by $\alpha'$ and $\beta'$).

**Proposition 5.5 (see [26, 34]).** *The Poisson structure on $G_-$ is defined by the following formula:*

$$(5.2) \qquad \eta_{G_-}(\alpha, \beta) = \left(\left(\mathcal{R}_g^* \alpha'\right)_+, \mathcal{R}_g^* \beta'\right) - \left(\left(\mathcal{L}_g^* \alpha'\right)_+, \mathcal{L}_g^* \beta'\right).$$

Here $\mathcal{R}_g^*, \mathcal{L}_g^* \colon T_g^* G \to T_e^* G = \mathfrak{G}^*$, $()_+$ is the projection of $\mathfrak{G}^*$ on $\mathfrak{G}_+^* = \mathfrak{G}_-$ along $\mathfrak{G}_-^*$, and $(,)$ is the pairing on $\mathfrak{G}^*$ corresponding to the invariant pairing $(,)$ on $\mathfrak{G}$.

*Proof.* It immediately follows from definition

$$\eta_{G_-}(\alpha, \beta) = \left\langle \left(\mathcal{R}_g^* \alpha'\right)_+, \left(\mathcal{R}_g^* \beta'\right)_- \right\rangle - \left\langle \left(\mathcal{L}_g^* \alpha'\right)_+, \left(\mathcal{L}_g^* \beta'\right)_- \right\rangle$$

and the isotropy property of $\mathfrak{G}_+$ and $\mathfrak{G}_-$.  □

*Important Remark.* It is easy to see that in order to define this Poisson structure on $G_-$ we do not need the entire group $G$, but only the tangent bundle of $G$ along $G_-$. To construct this bundle the existence of a group $G_+$ is not necessary, it is sufficient to know only the Lie algebra $\mathfrak{G}_+$ with the action of the group $G_-$ on $\mathfrak{G} = \mathfrak{G}_+ \oplus \mathfrak{G}_-$. Indeed, if $\alpha^\circ$ is any lifting of $\mathcal{R}_g^* \alpha \in T_e^* G_- = \mathfrak{G}_-^* = \mathfrak{G}^*/\mathfrak{G}_+^*$ to $\mathfrak{G}^*$, then the structure (5.2) can be rewritten as

$$(5.3) \qquad \eta_{G_-}(\alpha, \beta) = \left((\alpha^\circ)_+, \beta^\circ\right) - \left(\left(\operatorname{Ad}_g^* \alpha^\circ\right)_+, \operatorname{Ad}_g^* \beta^\circ\right).$$

*Remark* 5.6. One of the most interesting cases of Formula (5.2) deals with the group $G$ of invertible elements in an associative algebra $\mathfrak{A}$ and with a subgroup $G_-$ being an open subset of an affine subspace in $\mathfrak{A}$. The Poisson structure given by Formula (5.2) is quadratic in this case.

**5.2. Benney structure.** Here we discuss applications of the above formulae to the algebra of pseudodifferential symbols; in the next section we apply the formulas to its central extension.

Let $\mathfrak{g}$ be the Lie algebra of pseudodifferential symbols of integer order with the usual decomposition on differential $(+)$ and integral $(-)$ parts. Recall that the Lie group $G_- = G_{\text{int}}$ for the integral part $\mathfrak{g}_- = \mathfrak{g}_{\text{int}} \subset \mathfrak{g}$ consists of symbols $1 + P = 1 + \sum_{k=-\infty}^{-1} u_k(x) D^k$. Since $\mathfrak{g}_{\text{int}}$ is a Lie bialgebra (see Section 1), Formula (5.2) determines a Poisson structure on $G_{\text{int}}$. On the other hand, there is a remarkable structure on this set discovered in the theory of integrable systems.

**Definition 5.7 (see [22, 23, 24]).** The (second) Benney Poisson structure on $G_{\text{int}}$ is the structure defined by the following *Hamiltonian mapping* $\eta' \colon T_L^* G_{\text{int}} \to T_L G_{\text{int}}$. First, we represent the tangent vector from $T_L G_{\text{int}}$ as $\delta L \in \mathfrak{g}_{\text{int}}$:

$$L + \delta L = 1 + \sum_{k=-\infty}^{-1} (u_k(x) + \delta u_k(x)) D^k.$$

Secondly, we can represent a linear functional $\alpha \in T_L^* G_{\text{int}}$ by a differential operator $A_\alpha$ defined by $\langle \alpha, \delta L \rangle = \operatorname{Tr}(A_\alpha \circ \delta L)$. Now we can define $\eta'(\alpha)$ at $L$ as

$$(L \circ A_\alpha)_+ \circ L - L \circ (A_\alpha \circ L)_+ = -(L \circ A_\alpha)_- \circ L + L \circ (A_\alpha \circ L)_-.$$

*Remark* 5.8. The second part of the formula shows that this element is indeed an integral symbol, the first one shows that it is compatible with the fact that one can change $A_\alpha$ to $A_\alpha + B_\alpha$ (where $B_\alpha$ is an arbitrary integral symbol) without changing the resulting $\eta'(\alpha)$. (That means that $A_\alpha$ is in fact not a differential operator, but rather an element of the quotient of pseudodifferential symbols by integral ones.)

*Remark* 5.9. We have not proved that this Hamiltonian mapping satisfies the Jacobi identity, however, the following theorem implies that.

**Theorem 5.10.** *The Poisson structure $\eta_{G_{\text{int}}}$ given by Formula (5.2) coincides in the case of the algebra of integral (=Volterra) symbols with the Benney structure.*

*Proof.* First we note that the group

$$G_{\text{int}} = \{1 + P \mid P \text{ is an integral symbol}\}$$

is an affine subspace in $\mathfrak{g}$. To write down Formula (5.2) in this particular case let $\alpha$ be a cotangent vector to $G_-$ at $L \in G_-$. Representing a tangent vector as $L + \delta L$, $\delta L \in \mathfrak{g}_{\text{int}}$, we can consider $\alpha$ as a linear functional $\delta L \mapsto \langle \alpha, \delta L \rangle$ on the space of integral symbols. Identify it with some differential symbol $A$ using the pairing $\langle \alpha, \delta L \rangle = (A, \delta L) = \operatorname{Tr}(A \circ \delta L)$.

Now let the functional $\mathcal{R}_L^* \alpha$ corresponds in the same way to another differential operator $\bar{A}$. Let $B$ corresponds to $\beta$. A direct computation shows that:

$$\left\langle \bar{A}, \delta L \right\rangle = \left\langle \mathcal{R}_L^* \alpha, \delta L \right\rangle = \left\langle \alpha, \mathcal{R}_{L*} \left( \delta L \right) \right\rangle = \left\langle \alpha, \delta L \circ L \right\rangle = \left( A, \delta L \circ L \right) = \left( L \circ A, \delta L \right).$$

Hence $\bar{A}$ and $L \circ A$ have the same differential parts (i.e., $\left( \bar{A} \right)_+ = \left( L \circ A \right)_+$). Since the same is true for left translations, we can rewrite

$$\begin{aligned}
\eta_{G_-}|_L \left( \alpha, \beta \right) &= \left( \left( \mathcal{R}_L^* \alpha' \right)_+, \mathcal{R}_L^* \beta' \right) - \left( \left( \mathcal{L}_L^* \alpha' \right)_+, \mathcal{L}_L^* \beta' \right) \\
&= \left( \left( L \circ A \right)_+, L \circ B \right) - \left( \left( A \circ L \right)_+, B \circ L \right) \\
&= \left( \left( L \circ A \right)_+ \circ L - L \circ \left( A \circ L \right)_+, B \right).
\end{aligned}$$

Thus the Hamiltonian mapping $T_L^* G \to T_L G$ corresponding to the Poisson structure $\eta_{G_-}$ sends $A \mapsto \left( L \circ A \right)_+ \circ L - L \circ \left( A \circ L \right)_+$.   $\square$

## 6. Poisson—Lie geometry of the extended Lie group of pseudodifferential symbols

### 6.1. Gelfand—Dickey structure.
Now we extend the construction of the previous section to the algebra of pseudodifferential symbols enlarged by addition of the logarithm of the derivative. We start with definitions of the Gelfand—Dickey (also called Adler—Gelfand—Dickey or generalized $n$-KdV) structures.

In this section we finally drop the $\circ$ symbol for the product of pseudodifferential symbols, since it renders the formulae unreadable. As we promised, from now on we denote this product as $AB$, except in the cases when this can result in misunderstanding.

**Definition 6.1.** The (second generalized) Gelfand—Dickey Poisson structure on

$$\tilde{G}_{\text{int}} = \left\{ \left( 1 + \sum_{i=-\infty}^{-1} u_i \left( x \right) D^i \right) \circ D^t \right\}$$

is defined as follows:

(1) The value of the Poisson bracket of two functions at a given point is determined by the restriction of these functions to the subset $t = \text{const}$.

(2) The subset $t = \text{const}$ is an affine space, so we can identify its tangent space with the set of symbols of the form $\delta L = \left( \sum_{-\infty}^{-1} \delta u_i D^i \right) D^t$. We can also identify the cotangent space with the set of symbols of the form $A = D \, \bar{A}^{-t}$, where $\bar{A}$ is a differential operator, using the pairing

$$F_A \left( \delta L \right) \overset{\text{def}}{=} \left\langle A, \delta L \right\rangle = \text{Tr} \left( \delta L \circ A \right).$$

(3) Now it is sufficient to define the bracket on linear functionals $F_A$, and

$$\{F_A, F_B\}\,|_L = F_B\left(V_{F_A}\left(L\right)\right),$$

where $V_{F_A}\left(L\right)$ is the following Hamiltonian mapping $F_A \mapsto V_{F_A}\left(L\right)$ (from cotangent space $D^{-t} \circ \mathfrak{g}_{\mathrm{DO}}$ to the tangent one $\mathfrak{g}_{\mathrm{int}} \circ D^t$):

$$(6.1) \qquad\qquad V_{F_A}\left(L\right) = \left(LA\right)_+ L - L\left(AL\right)_+ .$$

*Remark* 6.2. Usually this definition is given only in the case when $t$ is a fixed positive integer and the symbol $L$ is a differential operator ([1, 12]). Another well known case is $t = 0$, when we get the Benney structure ([23, 22, 24]), and $t = 1$, when we get the KP structure (cf. [5]). Therefore this definition interpolates between the definitions of KP, Gelfand—Dickey and Benney structures.

**6.2. The key observation.**   It is the following fact that challenged us to write this paper:

**Theorem 6.3.** *The Poisson structure $\eta_{\widetilde{G}_{\mathrm{int}}}$ given by Formula (5.2) coincides in the case of the extended Lie group $\widetilde{G}_{\mathrm{int}}$ of classical pseudodifferential symbols with the second Gelfand—Dickey structure.*

*Proof.* Let $L \in \widetilde{G}_{\mathrm{int}}$, $L = (1 + P)\,D^t$, where $P$ is an integral symbol. Let $\alpha$ be a cotangent vector at the point $L$. We are going to use two different representations of this covector. First, we can write a tangent vector to $\widetilde{G}_{\mathrm{int}}$ as $(\delta P, \delta t)$, so we can find a number $\tau$ and a differential operator $A$ such that

$$\langle \alpha, (\delta P, \delta t)\rangle = (A, \delta P) + \tau \cdot \delta t \text{ for any } \delta t,\, \delta P.$$

Do the same for $\beta$ and $B$.

On the other hand, we can proceed in the same way as in Theorem 5.10. In this case $\mathfrak{g}_- = \widetilde{\mathfrak{g}}_{\mathrm{int}}$, $\mathfrak{g} = \widetilde{\mathfrak{g}}$. Consider the right translation $\mathcal{R}_L^* \alpha$ of $\alpha \in T_L^* \widetilde{G}_{\mathrm{int}}$ into $e \in \widetilde{G}_{\mathrm{int}}$. Let $\mathcal{R}_L^* \alpha \in T_e^* \widetilde{G}_{\mathrm{int}} = \widetilde{\mathfrak{g}}_{\mathrm{int}}^*$ corresponds in the same way to a pair $\left(\bar{A}, \bar{\tau}\right)$. To express $\bar{\tau}$ and $\bar{A}$ in terms of $\tau$ and $A$ we begin with the formula

$$\langle R_L^* \alpha, (\delta P, \delta t)\rangle = \langle \alpha, R_L\left(\delta P, \delta t\right)\rangle .$$

The left-hand side can be written as

$$\left(\bar{A}, \delta P\right) + \bar{\tau} \cdot \delta t.$$

The right-hand side is

$$\left\langle \alpha, (1 + \delta P)\,D^{\delta t}\left(1 + P\right) D^t - \left(1 + P\right) D^t\right\rangle \approx$$
$$\approx \left\langle \alpha, \left(\left(1 + P\right) D^{t+\delta t} - \left(1 + P\right) D^t\right) + \delta P\left(1 + P\right) D^t + \left[D^{\delta t}, \left(1 + P\right)\right] D^t\right\rangle$$
$$\approx \left(A, \delta P\left(1 + P\right) + \delta t\left[\log D, P\right]\right) + \tau \cdot \delta t.$$

Here $\approx$ denote equality modulo terms of order $o\,(dP, dt)$. Hence

$$\bar{A} = ((1+P)\,A)_+ \,, \ \bar{\tau} = \tau + (A, [\log D, P])\,.$$

The similar formulae can be obtained for the left translation:

$$\bar{A}_{\text{left}} = \left(D^{-t} A \,(1+P)\,D^t\right)_+ , \quad \bar{\tau}_{\text{left}} = \tau - (A, [\log D, P])\,.$$

Due to identity $\widetilde{\mathfrak{g}}^*_{\text{int}} = \widehat{\mathfrak{g}}_{\text{DO}}$ one can view the covector $\left(\bar{A}, \bar{\tau}\right)$ as consisting of a differential operator $\bar{A}$ and the central element with the coefficient $\bar{\tau}$.

Now we can apply Formula (5.3), denoting the pairing in $\widetilde{\mathfrak{g}}$ by $\langle,\rangle$:

$$\eta_{\widetilde{G}_{\text{int}}}|_L\,(\alpha, \beta) =$$
$$= \left\langle \left(((1+P)\,A)_+\,, \tau + (A, [\log D, P])\right), ((1+P)\,B, \tau + (B, [\log D, P]))\right\rangle$$
$$- \left\langle \left(\left(D^{-t} A\,(1+P)\,D^t\right)_+, \tau - (A, [\log D, P])\right), \left(D^{-t} B\,(1+P)\,D^t, \tau - (B, [\log D, P])\right)\right\rangle$$
$$= \left(((1+P)\,A)_+\,, (1+P)\,B\right) - \left(\left(D^{-t} A\,(1+P)\,D^t\right)_+, D^{-t} B\,(1+P)\,D^t\right).$$

(The terms containing the central element $(0, 1) \in \widetilde{\mathfrak{g}}_{\text{int}}$ do not matter in this formula since no term at any side of the inner product contains $\log D$ with non-zero coefficient.)

At this point what we have proved is compatible with the first condition of Definition 6.1:

**Claim 6.4.** *The function $t = \deg L$ of the degree of the $\Psi DS$ is the Casimir function, i.e.,*

$$\{\deg L, \varphi\,(L)\} = 0$$

*for any function $\varphi$ on $\widetilde{G}_{int}$.*

Indeed, the differential of the function $\deg L$ corresponds to the covector $(A, \tau) = (0, 1)$, and this covector goes to 0 under the Hamiltonian mapping of the Poisson structure $\eta$ by the above formula.

Hence we can restrict our attention to the level set of this function $\deg L \equiv t = $ const. Note that again this level set has a natural structure of an affine space: $L = (1+P)\,D^t$. This allows us to identify a cotangent vector to this subset at $L$ with the symbol $\tilde{A} = D^{-t}A$, i.e.,

$$\langle \alpha, \delta L \rangle = (A, \delta P) = \left(D^{-t}A, \delta P \cdot D^t\right) = \left(\tilde{A}, \delta L\right).$$

We extended here our pairing between integral symbols and differential operators to the pairing between symbols of the form $PD^t$ (with an integral symbol $P$) and symbols of the form $D^{-t}A$ (with a differential operator $A$). Now

$$\eta_{\widetilde{G}_{\mathrm{int}}}|_L(\alpha, \beta) = \left(\left((1+P)\,D^t\widetilde{A}\right)_+, (1+P)\,D^t\widetilde{B}\right) - \left(\left(\widetilde{A}\,(1+P)\,D^t\right)_+, \widetilde{B}\,(1+P)\,D^t\right)$$

$$= \left(\left(L\widetilde{A}\right)_+, L\widetilde{B}\right) - \left(\left(\widetilde{A}L\right)_+, \widetilde{B}L\right) = \left(\left(L\widetilde{A}\right)_+ L - L\left(\widetilde{A}L\right)_+, \widetilde{B}\right)$$

Hence the Hamiltonian mapping can be written as

$$\widetilde{A} \mapsto \left(L\widetilde{A}\right)_+ L - L\left(\widetilde{A}L\right)_+.$$

It is clear that

$$(6.4) \qquad \left(L\widetilde{A}\right)_- L - L\left(\widetilde{A}L\right)_- = -\left(\left(L\widetilde{A}\right)_+ L - L\left(\widetilde{A}L\right)_+\right),$$

since $L\widetilde{A}L - L\widetilde{A}L = 0$.  □

## 7. Applications to the KdV and KP hierarchies

### 7.1. From KP to KdV: a restriction to a Poisson submanifold.
What we described by now is the Poisson—Lie structure on the group of pseudodifferential symbols with the leading coefficient 1. However, anyone who have seen the formula for the Gelfand—Dickey Poisson structure on the set of *differential* operators with leading coefficient 1 can easily recognize Definition 6.1—the formulae are identical, only the underlying manifolds differ. The reason for this is the following:

**Proposition 7.1.** *The Poisson structure from Definition 6.1 can be restricted to the subset $\mathcal{D}_k$ of differential operators of order $k$ with the leading coefficient 1. This structure is quadratic in $L \in \mathcal{D}_k$ for $k \geq 2$, linear nonhomogeneous in $L \in \mathcal{D}_1$ for $k = 1$ in the matrix case, and with constant coefficients in $L \in \mathcal{D}_1$ for $k = 1$ in the scalar case.*

*Proof.* The restriction operation is well-defined if the submanifold is a union of symplectic leaves. So we need only to show that the image of the Hamiltonian mapping at any point of the submanifold consists of tangent vectors to the submanifold. This is clear from (6.1): if $L$ is a differential operator, then the right-hand side is also a differential operator.

Suppose $k = 1$. Then it is possible to extend a given covector on $\mathcal{D}_1$ to $\widetilde{G}_{\mathrm{int}}$ such that the corresponding $A$ is of $\deg A = 0$ (recall that in notations of the previous section we identify a covector with a symbol $\widetilde{A} = D^{-t}A$, $A \in \mathcal{D}$, and here $t = k = 1$). Hence $\deg L\widetilde{A} = \deg \widetilde{A}L = 0$ if $L \in \mathcal{D}_1$. Moreover, $\left(L\widetilde{A}\right)_+ = \left(\widetilde{A}L\right)_+ = \left(D\widetilde{A}\right)_+$, thus we can write the result of the Hamiltonian mapping as $\left[\left(D\widetilde{A}\right)_+, L\right]$. This shows that the bracket is linear nonhomogeneous. On the other hand, the variable part of $L$ is of degree 0, thus in the scalar case

$$V_{\widetilde{A}}(L) = \left[\left(D\widetilde{A}\right)_+, L\right] = \left[\left(D\widetilde{A}\right)_+, D\right] = -A'.$$

Here $'$ denotes $\frac{d}{dx}$ (recall that $A$ is a function). We see that in the affine coordinate system the Hamiltonian mapping is indeed translation invariant, since the right-hand side does not depend on $L$.

It is easy to see that there is no such cancellations for $k > 1$, thus the resulting structure is quadratic. $\square$

The possibility to restrict the Poisson structure to these submanifolds allows us to restrict various Hamiltonian systems from $\widetilde{G}_{\mathrm{int}}$ to $\mathcal{D}_k$. As we show it in Section 9, the standard Hamiltonian systems of $k$-KdV can be all obtained from a remarkable Hamiltonian system on $\widetilde{G}_{\mathrm{int}}$ of KP-type.

### 7.2. Poisson properties of the operator multiplication and Miura transform.
We are going to take dividends and deduce some properties of the Poisson bracket on the KdV manifold $\mathcal{D}_k$ that usually demand some ad hoc computation ([12, 35]).

**Corollary 7.2.** *The defined by the Hamiltonian mapping (6.1) bracket on $\mathcal{D}_k$*

$$\{f, g\}\,|_L = \langle \eta|_L, df|_L \wedge dg|_L \rangle$$

*satisfies the Jacobi identity.*

*Proof.* As we have seen, the corresponding bracket on $\widetilde{G}_{\mathrm{int}}$ satisfies the Jacobi identity (since it coincides with the Poisson—Lie bracket), therefore the restriction to differential operators also satisfies this identity. $\square$

**Corollary 7.3.** *The multiplication mapping*

$$\mathcal{D}_k \times \mathcal{D}_l \to \mathcal{D}_{k+l}$$

*is a mapping of Poisson manifolds $\mathcal{D}_k$, $\mathcal{D}_l$, and $\mathcal{D}_{k+l}$ equipped with the second Gelfand—Dickey Poisson structures.*

*Proof.* The multiplication mapping $\widetilde{G}_{\mathrm{int}} \times \widetilde{G}_{\mathrm{int}} \to \widetilde{G}_{\mathrm{int}}$ is a mapping of Poisson manifolds by definition of the Poisson—Lie structure. Taking the restriction to Poisson submanifolds $\mathcal{D}_k \times \mathcal{D}_l \subset \widetilde{G}_{\mathrm{int}} \times \widetilde{G}_{\mathrm{int}}$ and $\mathcal{D}_{k+l} \subset \widetilde{G}_{\mathrm{int}}$ proves the corollary. $\square$

**Corollary 7.4 (Kupershmidt—Wilson theorem [21]).** *Consider the Miura mapping (i.e., a mapping of multiplication of $k$ first order differential operators with scalar coefficients)*

$$\mathcal{D}_1^k \to \mathcal{D}_k.$$

*This is a fibration with a finite-dimensional fiber, and this mapping straightens the second Gelfand—Dickey Poisson structure, i.e., the constant Poisson bracket on the affine space $\mathcal{D}_1^k$ induced by the (constant) second Gelfand—Dickey bracket on $\mathcal{D}_1$ goes to the (quadratic) Poisson bracket on $\mathcal{D}_k$.*

*Proof.* Consider the $k$-dimensional space of (all) solutions $\operatorname{Ker} L$ of

$$L\varphi = 0$$

for a differential operator $L \in \mathcal{D}_k$. A factorization $L = L_1 L_2$ determines a subspace $\operatorname{Ker} L_2 \subset \operatorname{Ker} L$. It is easy to see that the operator $L$ is uniquely determined by $\operatorname{Ker} L$, therefore one can reconstruct the factorization of $L$ by the subspace $\operatorname{Ker} L_2$. Hence one can reconstruct a factorization

$$L = L_1 L_2 \ldots L_k, \quad L_i \in \mathcal{D}_1,$$

by the flag

$$\operatorname{Ker} L_k \subset \operatorname{Ker} L_{k-1} L_1 \subset \cdots \subset \operatorname{Ker} L_1 L_2 \ldots L_k = \operatorname{Ker} L.$$

This proves the fibration part of the corollary, because the corresponding flag space is finite-dimensional. To prove the second part we note that the second Gelfand—Dickey Poisson structure on $\mathcal{D}_1 = \{D_x + u(x)\}$ is constant in the standard coordinate system, while the multiplication mapping preserves the Poisson structures. $\quad\square$

*Remark* 7.5. Note that for $\alpha \notin \mathbb{Z}$ we do not have such a nice Poisson submanifold $\mathcal{D}_\alpha$. Moreover, it is not clear whether one can straighten "some part" of the Poisson submanifold $\{\deg g = \alpha\}$ by a mapping with a finite-dimensional fiber.

**7.3. The dressing action.** Let us consider now the corresponding version of the dressing action. Such an action is defined for any Poisson—Lie group.

Consider a left-invariant 1-form $\omega$ on a Poisson—Lie group $G$ with the tangent Lie bialgebra $(\mathfrak{G}, \mathfrak{G}^*)$. Let a vector field $V$ be the image of this 1-form under the *Hamiltonian mapping.* That means that if $df|_x = \omega_x$, then

$$\{f, g\}|_x = (V \cdot g)|_x$$

for any $g$, where $V \cdot g$ is the derivative of $g$ along $V$ (this condition determines $V$ uniquely). Taking into account that a left-invariant 1-form corresponds to a cotangent vector at $e \in G$, we get a mapping from $\mathfrak{G}^*$ into $\operatorname{Vect} G$. The fundamental fact [34] is that this mapping is a morphism of Lie algebras, so it defines a (local) action of the *dual group* $G^*$ (i.e., of a (local) Lie group with Lie algebra $\mathfrak{G}^*$) on $G$.

Moreover, the relation of $\mathfrak{G}$ and $\mathfrak{G}^*$ in the notion of a Lie bialgebra is absolutely symmetric, as the theorem on Manin triples shows. Therefore on the dual group there is a natural structure of a Poisson—Lie group, and the action of $G^*$ on $G$ is a Poisson—Lie action:

**Theorem 7.6** ([34, 26])**.** *Consider a Manin triple* $(\mathfrak{G}, \mathfrak{G}_-, \mathfrak{G}_+)$ *and the corresponding (local) Poisson—Lie groups* $G$, $G_-$ *and* $G_+$. *A vector* $X \in \mathfrak{G}_+$ *can be considered as a covector* $X \in \mathfrak{G}_-^*$. *Consider the corresponding left-invariant covector field* $\mathcal{L}(X) \in \Gamma(T^*G_-)$ *and the image of this field* $V(X)$ *under the Hamiltonian mapping on* $G_-$. *The mapping* $\mathfrak{G}_+ \to \mathrm{Vect}\, G_-$ *is a representation of* $\mathfrak{G}_+$, *and the corresponding (local) action of* $G_+$ *on* $G_-$ *is a Poisson—Lie action.*

**Definition 7.7.** This action is called a (left) dressing action. In the same way we can define a right dressing action and the corresponding notions for an action of a Lie bialgebra.

We cannot apply the discussed above method literally to the Poisson—Lie group $\widetilde{G}_{\mathrm{int}}$: there is no Lie group with Lie algebra $\mathfrak{g}_{\mathrm{DO}}$ of differential operators (or with its central extension $\widehat{\mathfrak{g}}_{\mathrm{DO}}$). However, we can do it if we stop when we have constructed the action of Lie algebra $\mathfrak{g}^*$ only. That means that in our case we get a Poisson—Lie action of the central extension $\widetilde{\mathfrak{g}}_{\mathrm{DO}}$ of the Lie algebra $\mathfrak{g}_{\mathrm{DO}}$ of differential operators on the Poisson manifold $\widetilde{G}_{\mathrm{int}}$. It is easy to see that the center acts trivially, therefore we get an action of $\mathfrak{g}_{\mathrm{DO}}$. Thus

**Corollary 7.8.** *There is a Poisson—Lie action of the Lie bialgebra of differential operators on the group of pseudodifferential symbols* $\widetilde{G}_{\mathrm{int}}$.

On the other hand, A. Radul defined in [29] an action of this Lie algebra on the Poisson submanifold $\mathcal{D}_k$ of differential operators. Without any surprise, the latter action is a restriction of the former from $\widetilde{G}_{\mathrm{int}}$ to $\mathcal{D}_k$, as shows the following

**Corollary 7.9.** *The dressing action of the Lie bialgebra of differential operators can be restricted to the subset* $\mathcal{D}_k \subset \widetilde{G}_{\mathrm{int}}$. *If* $L \in \mathcal{D}_k$, *and* $A$ *is a differential operator, then the corresponding to* $-A$ *tangent vector to* $\mathcal{D}_k$ *at* $L$ *can be described as*

$$(-A) \cdot L = LA - \left(LAL^{-1}\right)_+ L = \left(LAL^{-1}\right)_- L,$$

*i.e., it is the remainder in the right division of* $LA$ *by* $L$.

*Proof.* We can write the left-invariant 1-form on $\widetilde{G}_{\mathrm{int}}$ that corresponds to $-A$ as

$$\delta L|_L \mapsto -\left\langle L^{-1}\delta L, A \right\rangle = -\mathrm{Tr}\left(\delta L \cdot AL^{-1}\right).$$

Thus the pseudodifferential symbol that corresponds to this covector at $L$ is $-AL^{-1}$. The image of this covector under the Hamiltonian mapping is

$$-\left(L\left(AL^{-1}\right)\right)_+ L + L\left(\left(AL^{-1}\right)L\right)_+ = LA - \left(LAL^{-1}\right)_+ L.$$

□

*Remark* 7.10. Note that the formula in the corollary (but not the interpretation as a remainder) remains valid for an arbitrary $L$.

## 8. Relation of the GL− and SL− Poisson structures

**8.1. A quotient by the action of functions.**  What we have described is an analogue of the $GL_n$-Gelfand—Dickey structure. In this section we describe an analogue of the $SL_n$-case, i.e., the Poisson structure on the set of pseudodifferential symbols of the form

$$\left(1 + u_{-2}\left(x\right)D^{-2} + u_{-3}\left(x\right)D^{-3} + \dots\right)D^t.$$

We work mainly in the case of scalar coefficients, however, the case of matrix coefficients needs only cosmetic changes in the hypotheses of the theorems.

**Theorem 8.1.** *Consider the quotient of the adjoint (=dressing) action of functions (i.e., of differential operators of zeroth order) on $\widetilde{G}_{\text{int}}$ (in the case of scalar coefficients). It carries a Poisson structure, thus determines a Poisson structure on the space of pseudodifferential symbols $\{(1 + u_{-2}D^{-2} + u_{-3}D^{-3} + \dots)D^t\}$ without subleading term. For a fixed $t = n \in \mathbb{N}$ the subset of differential operators of such a form is a Poisson submanifold, the Poisson structure on it coincides with the second $SL_n$—Gelfand—Dickey structure.*

*Proof.* The proof of this theorem occupies the rest of this section. To identify this submanifold with a subspace in the *quotient* or (what is the same in this particular case, as we show in the following section) with a *Poisson reduction* of the GL-structure by the action of scalar functions (i.e., of differential operators of order 0), we need a notion of Poisson—Lie subgroups from Section 1. Furthermore, we need a description when a dressing action is a Hamiltonian action.

**Proposition 8.2.** *Let a Poisson—Lie group $G$ with Lie algebra $\mathfrak{G}$ acts on a Poisson manifold $X$ in a Poisson—Lie way, and let $H$ be a connected subgroup of $G$ with Lie algebra $\mathfrak{h} \subset \mathfrak{G}$ such that the orthogonal complement $\mathfrak{h}^\perp$ to $\mathfrak{h}$ in $\mathfrak{G}^*$ is a subalgebra. Then there is a natural Poisson structure on $H \backslash X$ such that the mapping $X \to H \backslash X$ is a mapping of Poisson manifolds. Moreover, if $H$ is a Poisson—Lie subgroup, i.e., $\mathfrak{h}^\perp$ is an ideal in $\mathfrak{G}^*$, then the action of $H$ on $X$ is a Poisson—Lie action.*

*Remark* 8.3. One can extend this proposition to the case when $H$ is a coisotropic submanifold of $G$, i.e., a Poisson bracket of two vanishing on $H$ functions vanishes on $H$. This condition is equivalent to the above condition in the case when $H$ is connected, as shows a simple modification of the proof of Lemma 1.12.

**Lemma 8.4.** *Consider a Lie bialgebra $(\mathfrak{G}, \mathfrak{G}^*)$. Then the abelian quotient $\mathfrak{a} = \mathfrak{G}/[\mathfrak{G}, \mathfrak{G}]$ of $\mathfrak{G}$ has a natural Lie bialgebra structure.*

*Proof.* Let $\mathfrak{i} = [\mathfrak{G}, \mathfrak{G}]$. It is sufficient to show that $\mathfrak{a}^* = \mathfrak{i}^\perp \subset \mathfrak{G}^*$ is a Lie subalgebra of $\mathfrak{G}^*$. Consider a Manin triple $\left(\bar{\mathfrak{G}}, \mathfrak{G}, \mathfrak{G}^*\right)$. An element $a \in \mathfrak{G}^* \subset \bar{\mathfrak{G}}$ is in $\mathfrak{i}^\perp$ iff $(X, a) = 0$ provided $X \in [\mathfrak{G}, \mathfrak{G}] \subset \mathfrak{G} \subset \bar{\mathfrak{G}}$. Now

$$([X_1, X_2], a) = (X_1, [X_2, a]),$$

therefore $a \in \mathfrak{i}^\perp$ iff $[a, \mathfrak{G}] \subset \mathfrak{G}$. Since the normalizer of the subalgebra $\mathfrak{G}$ in $\bar{\mathfrak{G}}$ is a Lie subalgebra, $\mathfrak{i}^\perp$ is a Lie subalgebra. $\quad\square$

*Remark* 8.5. This lemma has a direct counterpart in the theory of Hopf algebras: the quotient of a Hopf algebra by the two-sided ideal generated by commutators is a (commutative) Hopf algebra. Thus basing on a Hopf algebra we can construct two group schemes (i.e., commutative Hopf algebras) starting from the given Hopf algebra and from its dual.

**Corollary 8.6.** *Consider a connected simply connected Poisson—Lie group $G$ and the (say, left) dressing action of the dual Lie algebra $\mathfrak{G}^*$ on it. Consider the subspace $V$ in $\mathfrak{G}^*$ formed by fixed vectors in the coadjoint representation of $G$. Then*

(1) *this subspace is a Poisson—Lie subalgebra in $\mathfrak{G}^*$ with a trivial Lie algebra structure on $V^*$;*
(2) *the dressing action of this Lie subalgebra on $G$ is Hamiltonian;*
(3) *the Hamiltonian $H_\alpha$ corresponding to an element $\alpha \in V$ is a Casimir function iff $\alpha \in \text{Cent}\,\mathfrak{G}^*$.*

*Proof.* The part concerning the Poisson—Lie properties of $V$ is already proven since $V = \mathfrak{i}^\perp$ in the notations of the previous lemma. To proof the claim on the existence of Hamiltonians we note first that the dressing action of elements of $V$ preserves the Poisson structure, since this action is a Poisson—Lie action and the Lie algebra structure on the dual space $V^*$ is abelian. Moreover, by definition the dressing action is a result of the composition

$$\mathfrak{G}^* \xrightarrow{\lambda} \Omega^1(G) \to \text{Vect}(G),$$

where $\Omega^1(G)$ is equipped with the natural Lie algebra structure associated with the Poisson structure on $G$, the first arrow is the isomorphism to the subspace of left-invariant forms, the right arrow is the Hamiltonian mapping (both arrows are Lie algebra morphisms).

Let us show that the image of $V$ in $\Omega^1(G)$ consists of closed forms. Indeed, by the geometrical description of the cochain complex of a Lie algebra, if $\alpha \in \mathfrak{G}^* = C^1(\mathfrak{G})$, then $d\lambda(\alpha) \in \Omega^2(G)$ is a left-invariant form that corresponds to the element $d\alpha \in \Lambda^2\mathfrak{G}^* = C^2(\mathfrak{G})$. However, $V = Z^1(\mathfrak{G})$, since the condition

$$\langle d\alpha, X \wedge Y \rangle = \langle \alpha, [X, Y] \rangle = -\langle \text{ad}_X^* \alpha, Y \rangle = 0 \quad \text{for any } X, Y \in \mathfrak{G}$$

is exactly the condition on the subspace $V$. Now let us associate to an element $\alpha \in V$ the Hamiltonian $H_\alpha$ that is the only primitive of the closed form $\lambda(\alpha)$ that vanishes at $e \in G$. By definition the dressing action of $\alpha$ is exactly the Hamiltonian flow of $H_\alpha$. What remains to prove is the identity

$$H_{[\alpha,\beta]} = \{H_\alpha, H_\beta\}.$$

However, the fact that $\lambda$ is a morphism of Lie algebras, and the compatibility of $d \colon \Omega^0 \to \Omega^1$ with Poisson bracket on $\Omega^0$ and the bracket on $\Omega^2$ imply that the difference of the sides of this would-be-equality is a constant. This constant is equal to zero since the left-hand side vanishes at $e \in G$ by definition of $H_\bullet$, and so does the right-hand side due to the vanishing Poisson bracket at $e \in G$.

If $H_\alpha$ is a Casimir function, then it is constant on any symplectic leaf in $G$, i.e., on any orbit of the dressing action [34], hence $\alpha \in \operatorname{Cent} \mathfrak{G}^*$. (Another proof of this fact is a direct application of Formula (5.2).)    $\square$

*Remark* 8.7. It is easy to understand that the above Hamiltonians are characters of 1-dimensional representations of $G$.[6] Below, in Section 9, we discuss the Poisson—Lie properties of characters of arbitrary representations, i.e., invariant functions on $G$.

**8.2. The case of $\widetilde{G}_{\mathbf{int}}$.**   Consider the (abelian in the scalar case) Lie algebra $\mathfrak{o}$ consisting of multiplications by functions. Consider it as a subalgebra in the Lie algebra $\mathfrak{g}_{DO}$ of differential operators. Its orthogonal complement in $\mathfrak{g}_{DO}^* = \mathfrak{g}_{int}$ consists clearly of pseudodifferential symbols of degree $\leq -2$, i.e., it is an ideal in $\mathfrak{g}_{int}$. This remains true if we consider 1-dimensional extensions $\widehat{\mathfrak{o}}$, $\widehat{\mathfrak{g}}_{DO}$, $\widetilde{\mathfrak{g}}_{int}$ of $\mathfrak{o}$, $\mathfrak{g}_{DO}$ and $\mathfrak{g}_{int}$. Therefore there is a natural Poisson structure on the *quotient* $\mathcal{M}/\widehat{\mathfrak{o}}$ if $\widehat{\mathfrak{g}}_{DO}$ acts on $\mathcal{M}$ in a Poisson—Lie way.

*Remark* 8.8. Since the center of the Lie algebra $\widehat{\mathfrak{o}}$ acts trivially on $\widetilde{G}_{int}$, below we freely mix the Lie algebras $\mathfrak{o}$ and $\widehat{\mathfrak{o}}$, and Lie groups $O$ and $\widehat{O}$ in our discussion.

Consider the quotient $\mathcal{M}/\widehat{\mathfrak{o}}$ in the case $\mathcal{M} = \widetilde{G}_{int}$, or, better, in the case of a submanifold $\mathcal{M} = \{\deg = \operatorname{const} \neq 0\}$ in $\widetilde{G}_{int}$. We have seen (compare Corollary 7.9) that the action of $f \in \mathfrak{o}$ adds to a pseudodifferential symbol

$$L = \left(1 + u_{-1}(x) D^{-1} + \dots\right) D^t$$

a term $\varepsilon[f, L]$. (Here $\varepsilon$ in an infinitesimal parameter.) Therefore this action integrates to the action of the Lie group $O$ of invertible functions (of the form

$$O = \{F \mid F(x) > 0 \text{ for all } x\}$$

---

[6]Note that the tensor product of two such representations is again 1-dimensional. This is a key reason why such functions should generate a Hopf subalgebra.

in the real case and with the winding number 0 in the complex case) by the rule

$$F \cdot L = F \circ L \circ F^{-1}.$$

Thus the action on the coefficient $u_{-1}(x)$ is $u_{-1} \to u_{-1} - t \left(\log F(x)\right)'$, $t = \deg L$. It is easy to see that for $t \neq 0$ any orbit contains only one point with $u_{-1}(x) = \text{const}$, therefore we can identify the quotient with the manifold

$$\bar{\mathcal{M}} = \left\{ \left(1 + \phi D^{-1} + u_{-2}(x) D^{-2} + \dots\right) D^t \right\} \subset \mathcal{M}, \ \phi \text{ is a number.}$$

Actually, this constant $\phi$ is the average value of the original coefficient $u_{-1}(x)$.

**Lemma 8.9.** *The average value $\phi$ is a Casimir function on the space $\mathcal{M}$, i.e., $\{\phi, \varphi\} = 0$ for any function $\varphi$ on $\mathcal{M}$.*

*Proof.* Indeed, consider $[\widetilde{\mathfrak{g}}_{\text{int}}, \widetilde{\mathfrak{g}}_{\text{int}}]$. The orthogonal complement to this subspace of $\widetilde{\mathfrak{g}}_{\text{int}}$ is the subset of $\widehat{\mathfrak{g}}_{\text{DO}}$ consisting of "connections" (with a coefficient)

$$(\mu D + u(x), \gamma),$$

here $\mu$ and $\gamma$ are numbers. The intersection of the center of $\widehat{\mathfrak{g}}_{\text{DO}}$ with this set is spanned by the central element (0,1) and by (1,0), i.e., by $u(x) \equiv 1$. The above lemmas show that the corresponding Casimir functions on $G$ are deg and $\phi$ respectively. Moreover, an arbitrary function $u(x)$ corresponds to a Hamiltonian which sends

$$L = \left(1 + u_{-1}(x) D^{-1} + u_{-2}(x) D^{-2} + \dots\right) \circ D^t \mapsto \int u_{-1}(x) u(x) \, dx,$$

and the connection $D$ corresponds to the Hamiltonian

$$L = \left(1 + u_{-1}(x) D^{-1} + u_{-2}(x) D^{-2} + \dots\right) \circ D^t \mapsto \int u_{-2}(x) \, dx.$$

$\square$

Thus

(1) the submanifold

$$\mathcal{M}^s = \{L \mid \phi(L) = s\} \subset \mathcal{M}$$

   is a Poisson submanifold, and we can restrict our attention to the submanifold $\mathcal{M}^0$;

(2) any orbit of the $O$-action on the submanifold $\mathcal{M}^0$ contains only one point on $\mathcal{M}^0 \cap \bar{\mathcal{M}}$ (i.e., with $u_{-1} = 0$), therefore one can identify $\mathcal{M}^0 / O$ with $\mathcal{M}^0 \cap \bar{\mathcal{M}} \overset{\text{def}}{=} \bar{\mathcal{M}}^0$.

Consider the corresponding quotient Poisson structure on this manifold $\bar{\mathcal{M}}^0$. The definition of the Poisson structure on the quotient gives the following algorithm: To compute the Hamiltonian mapping on a 1-form on $\bar{\mathcal{M}}^0$ we should extend this 1-form to $\mathcal{M}$ in such a way that it vanishes on any tangent to an $O$-orbit vector, and compute the Hamiltonian mapping on $\mathcal{M}$. The result is a vector field on $\mathcal{M}$. To get a vector field on $\bar{\mathcal{M}}$ we have to take its projection on $\bar{\mathcal{M}}^0$ along the orbits of $O$.

Launching into calculations let $L \in \bar{\mathcal{M}}$, and let $\alpha$ be a covector in $T_L^* \bar{\mathcal{M}}$. Recall that $\bar{\mathcal{M}}$ is an affine subspace in the affine space $\mathcal{M}$, and that we have identification of the dual vector space to $\mathcal{M}$ with the set of pseudodifferential symbols of degree $-t$.

The cotangent vector space to $\bar{\mathcal{M}}$ is a quotient-space of $T_L^* \mathcal{M}$, and we need to lift $\alpha$ to get an element of the latter space. The above condition on this lifting $\tilde{\alpha}$ is that $\tilde{\alpha}$ is orthogonal to $[\mathfrak{o}, L]$:

$$\langle \tilde{\alpha}, [f, L] \rangle = 0 \text{ for all } f \in O.$$

As usual, we identify $\tilde{\alpha}$ with a pseudodifferential symbol $\tilde{A}$ of the form $D^{-t}A$, $A$ being a differential operator $A = \sum_{i=0}^n v_i(x) D^i$:

$$\langle \tilde{\alpha}, \delta L \rangle = \mathrm{Tr}\left( \tilde{A} \circ \delta L \right), \quad \deg \delta L \leq t - 1.$$

Then the condition above becomes $\mathrm{Tr}\left( \left[ \tilde{A}, L \right] \circ f \right) = 0$, and thus $\mathrm{res}\left[ \tilde{A}, L \right] \equiv 0$. That means that the lowest-order term $D^{-t}v_0(x)$ of $\tilde{A}$ is uniquely determined by the others (up to an additive constant). Let us identify the dual vector space to $\bar{\mathcal{M}}$ with the space of symbols with a constant lowest-order term $v_0$. We can write this identification explicitly as

$$\tilde{A} \mapsto \bar{A} \stackrel{\mathrm{def}}{=} \tilde{A} - D^{-t} \circ \left( \int (\text{the coefficient at } D^{-1} \text{ in } \left[ \tilde{A}, L \right]) \, dx \right).$$

Finally we apply the Hamiltonian mapping (6.1) to $\bar{A}$. It is easy to see that the difference $\bar{A} - \tilde{A}$ adds to the action of $\tilde{A}$ a term of the form $[f, L]$, where $f$ is a function. Therefore this difference is being killed anyway by taking the quotient by the action of $O$. On the other hand, the coefficient at $D^{-1}$ in $\left[ \bar{A}, L \right]$ is 0, therefore the corresponding to $\bar{A}$ vector is already tangent to $\bar{\mathcal{M}}$. That means that we get the usual description of the second Gelfand—Dickey Poisson structure on the set of (pseudo)differential symbols without subleading term as a quotient by the action of functions: apply (6.1), but substitute such an $\tilde{A}$ (from the set of differential operators corresponding to the same covector) that the result has a zero coefficient at $D^{t-1}$.   $\square$

*Remark* 8.10. In fact we can describe the above manifold as a surface of codim $= 2$ in the quotient of $\tilde{G}_{\mathrm{int}}$ by the action of functions. Indeed, the functions deg and $\phi$ can be pushed down to the quotient, and the level sets of these functions form a foliation of codimension 2. Above we worked with one leaf of this foliation. In the following sections we are going to describe how to interpret these two steps (taking the quotient and a subset) as a result of the Poisson reduction.

**8.3. A Poisson reduction.**    The usual notion in the Hamiltonian approach to the classical mechanics is the *symplectic reduction*. Once we know one conserved quantity in the dynamics of a Hamiltonian system we can reduce the dimension of the system by two. Several quantities that are *in involution* make it possible to reduce the number of degrees of freedom by twice their number. Roughly speaking, the symplectic reduction is the description of what is possible to do without the requirement that the quantities are in involution (and the *Poisson reduction* is the generalization on the Poisson case).

In that case one adds to the set of the quantities all the functions on the phase space that can be expressed via the Poisson bracket and the above quantities. After this procedure the obtained set of functions is closed with respect to the Poisson bracket. Denote this Lie algebra as $\mathfrak{A}$. This set of quantities can be described as a Lie algebra $\mathfrak{A}$ and a morphism $\mathfrak{A} \overset{i}{\hookrightarrow} \mathrm{Func}\,(\mathcal{M})$ of inclusion of it into the Poisson algebra of the phase space $\mathcal{M}$. Given a point $m \in \mathcal{M}$ one constructs a linear function

$$\mu\,(m) \overset{\mathrm{def}}{=} (\mathfrak{A} \ni f \mapsto i\,(f)\,|_m)$$

on $\mathfrak{A}$, i.e., one gets a mapping $\mu\colon \mathcal{M} \to \mathfrak{A}^*$.

**Definition 8.11.** This mapping is called the *momentum mapping*.

A simple check shows that this mapping is compatible with the Poisson structure on $\mathcal{M}$ and the Lie—Beresin—Kirillov—Kostant structure on $\mathfrak{A}^*$. Moreover, the ad*-action of $\mathfrak{A}$ on $\mathfrak{A}^*$ is compatible with the natural action of $\mathfrak{A}$ on $\mathcal{M}$ (the action of a function $f \in \mathfrak{A}$ is the Hamiltonian vector field corresponding to this function).

**Definition 8.12.** Fix an ad*-orbit $\mathcal{O}$ in $\mathfrak{A}^*$. Then the *Poisson reduction* of $\mathcal{M}$ with respect to $\mathcal{O}$ is the object described in either one of the following two ways:

(1) it is $\mu^{-1}\,(\mathcal{O})\,/\mathfrak{A}$;
(2) it is the preimage of $\mathcal{O}/\,\mathrm{ad}_{\mathfrak{A}}^*$ in $\mathcal{M}/\mathfrak{A}$.

Here the quotient by the action of a Lie algebra means the same as the quotient by the action of the corresponding group.

*Remark* 8.13. Starting with one Hamiltonian $P$ (that automatically commutes with itself), we obtain the usual reduction by two dimensions: $\dim \mathfrak{A} = \dim \mathfrak{A}^* = 1$, $\mathrm{codim}\,\mathcal{O} = 1$, therefore $\mu^{-1}\,(\mathcal{O})$ is of codimension one and taking the quotient by the action of $\mathfrak{A}$ on $\mu^{-1}\,(\mathcal{O})$ kills one more dimension.

**Example 8.14.** Starting with two conjugate coordinates $P$ and $Q$, we obtain a three-dimensional Heisenberg algebra $\mathfrak{A} = \langle P, Q, c \rangle$ with $c = 1$ on $\mathcal{M}$ (since $\{P, Q\} = 1$). The Poisson reduction with respect to $\mathcal{O} \subset \mathfrak{A}^*$ is nontrivial only if $c|_{\mathcal{O}} = 1$ (there is exactly one such orbit), and it coincides with the Poisson reduction with respect to the action of any one of $P$ or $Q$. We see that if the Hamiltonians are not in involution, it is not so simple to use them for reduction of dimension.

*Remark* 8.15. Suppose we start with an action of a Lie algebra $\mathfrak{G}$ on a Poisson manifold $\mathcal{M}$ that preserves the Poisson structure and symplectic leaves in $\mathcal{M}$. For any element $X \in \mathfrak{G}$ we can find (locally) a Hamiltonian $H_X$ on $\mathcal{M}$ that defines the same vector field on $\mathcal{M}$ as $X$. However, nothing guaranties that $H_{[X,Y]} = \{H_X, H_Y\}$, these two functions can differ by a Casimir function on $\mathcal{M}$. This means that to get a Lie subalgebra in functions, we need to consider *all* Hamiltonians that correspond to a given element $X \in \mathfrak{G}$, resulting in a central extension $\widehat{\mathfrak{G}}$ of $\mathfrak{G}$ by the space of Casimir functions on $\mathcal{M}$. After this the above construction can be carried out. In other words, *an action of $\mathfrak{G}$ that preserves the Poisson structure on $M$ leads to a canonically defined central extension $\widehat{\mathfrak{G}}$ and a mapping $M \to \widehat{\mathfrak{G}}^*$. One calls these data *the momentum mapping for the action of $\mathfrak{G}$ on $M$.*

**8.4. The result of the Poisson reduction.** It turns out that we work in a very rare situation when the Poisson reduction almost coincides with taking a quotient:

**Proposition 8.16.** *The second Gelfand—Dickey structures on the sets*

  (1) *of classical pseudodifferential symbols with the leading coefficient one without the subleading term;*
  (2) *and of differential operators with the leading coefficient one without the subleading term;*

*are results of a Poisson reduction by the dressing (=adjoint) action of centrally extended invertible functions on the Poisson—Lie group $\widetilde{G}_{int}$ of classical pseudodifferential symbols and on the closed subset of differential operators respectively.*

*Proof.* To compute the Poisson structure on the quotient $\widetilde{G}_{int}/\widehat{O}$ we apply the fact that $\widehat{\mathfrak{o}}$ is a subset of $[\widetilde{\mathfrak{g}}_{int}, \widetilde{\mathfrak{g}}_{int}]^{\perp}$. Therefore the dressing action of this Lie algebra is Hamiltonian. Consider the corresponding momentum mapping $\widetilde{G}_{int} \xrightarrow{\mu} \widehat{\mathfrak{o}}^*$. The discussion in Section 8.1 shows that the central charge of $\mu(L)$ is equal to $\deg L$. This provides one Casimir function on $\widehat{\mathfrak{o}}^*$. On the open subset in $\widehat{\mathfrak{o}}^*$ where the central charge is non-zero the symplectic leaves are of codimension two, and we know that another Casimir function on $\widehat{\mathfrak{o}}^*$ corresponds to the function $\phi$ on $\widetilde{G}_{int}$. Moreover, a stabilizer of an element in this open subset coincides with the center of $\widehat{\mathfrak{o}}$, which acts trivially on $G_{int}$. Therefore the Poisson reduction on the preimage $\widetilde{G}_{int} \smallsetminus G_{int}$ of this open subset coincides with taking the quotient by the action of $\widehat{O}$ on the subset $\{\deg = \mathrm{const}, \phi = \mathrm{const}\}$. And this exactly what we did in Section 8.1.  $\square$

*Remark* 8.17. Consider the restriction to the hyperplane $t = 1$. The KP hierarchy of equations is usually considered on the space of symbols of the form

$$D + u_{-1}(x) D^{-1} + u_{-2} D^{-2} + \dots.$$

We see that this is exactly the result of Poisson reduction with respect to the action of $\widehat{O}$ (or $O$) on $\widetilde{G}_{\text{int}}$. The corresponding orbit in $\widehat{\mathfrak{o}}^*$ is the orbit containing the Maurer—Cartan 2-cocycle[7]

$$c\left(f, g\right) = \text{Tr}\left(\left[\log D, f\right] \circ g\right) = \int f \, dg.$$

This shows that the KP equation is an evolution equation (=ODE) on a Poisson submanifold of a quotient $\widetilde{G}_{\text{int}}/\operatorname{Ad}_O$ of a Poisson—Lie group $\widetilde{G}_{\text{int}}$ by the adjoint action of the group of functions with respect to multiplication. It is known also that the usual KP hierarchy can be extended to the space of symbols of the form

$$D + u_0\left(x\right) + u_{-1}\left(x\right) D^{-1} + u\left(x\right)_{-2} D^{-2} + \dots$$

in such a way that all the formulae remain the same, only the underlying manifold changes. Therefore one can consider the whole hypersurface $t = 1$ in $\widetilde{G}_{\text{int}}$ as a phase space of the (extended) KP hierarchy.

In the following section we suggest a Poisson—Lie-group-theoretic approach to the Hamiltonians that determine the KP hierarchy. This point of view on the KP hierarchy is elaborated in [16], were a self-consistent $q$-deformation is defined.

*Remark* 8.18. The same theory as above can be applied in the case of matrix differential and matrix pseudodifferential symbols. The only significant change is the description of the quotient of $\widetilde{G}_{\text{int}}$ by the adjoint action of matrix-valued functions. Since the adjoint action of $f$ on $L = \left(1 + u_0\left(x\right) D^{-1} + \dots\right) D^t$ and on $\left(L^{1/t}\right)_+ = D + \frac{1}{t} u_0\left(x\right)$ are related by the $t$-th power, it is clear that the "invariant part" of the coefficient $u_0$ is the monodromy of the connection $\left(L^{1/t}\right)_+$. On the other hand, we can reduce a generic connection to a connection with constant diagonal coefficients by a conjugation, and the function we conjugate with is defined up to a multiplication by a constant diagonal matrix.

So an open subset in the quotient can be described as the set of operators

$$L = \left(1 + \phi D^{-1} + u_1\left(x\right) D^{-2} + \dots\right) D^t.$$

Here $\phi$ is a constant diagonal matrix, $u_1\left(x\right)$ is defined up to a conjugation by a constant diagonal matrix.

*Remark* 8.19. The above arguments rely on the fact that $\widehat{\mathfrak{o}}$ is a subset of $\left[\widetilde{\mathfrak{g}}_{\text{int}}, \widetilde{\mathfrak{g}}_{\text{int}}\right]^{\perp}$. The latter set contains also an additional element $D$. The Hamiltonian for $D$ is the average value $\bar{u}_{-2}$ of the coefficient $u_{-2}\left(x\right)$ of the operator $L$ in the standard notation. Therefore the reduction consists of consideration of the level set $\bar{u}_{-2} = \text{const}$ and the quotient by the action of translations. Say, one can fix a section in the latter quotient considering $u_{-2}\left(0\right) = 0$ or fixing a phase of the first Fourier coefficient of $u_{-2}$.

---

[7]The 2-cocycle which determines a central extension is naturally identified with a vector in the dual space to this extension.

*Remark* 8.20. Let us investigate the behavior of the dressing (=Radul) action under the above reduction. We know that once we fix the average value of the coefficient $u_{-1}(x)$, the reduction is just a quotient by the (adjoint=dressing) action of functions on the circle. Thus the problem is reduced to the following: consider the action of a Lie algebra $\mathfrak{G}$ on $\mathcal{M}$ and take the quotient by the action of a Lie subalgebra $\mathfrak{H} \subset \mathfrak{G}$. What remains of the action of $\mathfrak{G}$ on $\mathcal{M}/\mathfrak{H}$?

The natural answer is that on $\mathcal{M}/\mathfrak{H}$ acts the (Weyl) algebra Norm $(\mathfrak{H})/\mathfrak{H}$. Look what does it mean in our case, where $\mathfrak{G} = \mathfrak{g}_{\mathrm{DO}}$, $\mathfrak{H} = \mathfrak{o}$. Evidently, Norm $(\mathfrak{o})$ is the Lie algebra $\mathcal{D}^{(\leq 1)}$ of differential operators of order $\leq 1$. Thus the Weyl algebra in this case is the Lie algebra of vector fields. On the other hand, T. Khovanova described in [19] an action of the Virasoro algebra on Gelfand—Dickey manifold. It is easy to see that this action is the above action of Vect $(S^1)$ on the Gelfand—Dickey manifold.

**Corollary 8.21.** *The Khovanova action is a result of Poisson reduction applied to the Radul action.*

*Remark* 8.22. It is interesting to describe a Poisson version of the above picture that includes the central charge for Virasoro in it.

## 9. KP Hamiltonians on the Poisson—Lie group

We have seen that the Poisson submanifold $t = 1$ of the Poisson—Lie group $\widetilde{G}_{\mathrm{int}}$ is the usual KP manifold with the second Gelfand—Dickey Poisson structure. However, in the theory of integrable systems the Hamiltonians are no less important than the Poisson structures themselves.

In this section we introduce a hierarchy of Hamiltonian equations on the Poisson—Lie group of pseudodifferential symbols. The corresponding Hamiltonians satisfy simple commutation relations: they are divided into two natural groups, the Hamiltonians in the first group $\{K_\varphi, c\}$ commute as the elements of a Heisenberg algebra, the elements of the second group $\{H_i\}$ are central. The usual (=reduced) KP hierarchy is the result of the Poisson reduction with respect to the first group of Hamiltonians, i.e., the result of restrictions $K_\varphi = 0$, $c = 1$. After this reduction the Hamiltonians $H_i$ become the standard local Hamiltonians of the KP theory.

Moreover, after the reduction $K_\varphi = 0$, $c = 0$ we get the usual Hamiltonians and the Poisson structure of the Benney hierarchy, and the $n$-KdV-hierarchy is the restriction to a Poisson subset of the reduction $K_\varphi = 0$, $c = n$.

Recall that the usual Hamiltonians in the KP theory are

$$\widetilde{H}_k \colon L = D + u_0(x) + u_{-1}(x) D^{-1} + u_{-2}(x) D^{-2} + \cdots \mapsto \operatorname{Tr} L^k, \qquad k = 1, \ldots.$$

In the standard form of KP equation $u_0 = 0$, however, we postpone this reduction until later. The principal property $\left\{ \widetilde{H}_k, \widetilde{H}_l \right\} = 0$ remains true without this restriction.

**9.1. Invariant functions are closed under Poisson bracket.**   It is known that the $\mathrm{Ad}_G$-invariant functions on a Poisson—Lie group $G$ are in involution if the Poisson structure on $G$ is associated with an $r$-matrix $r \in \Lambda^2 \mathfrak{G}$ of the form (5.1) (see [33]). Here we recall what remains of this property for arbitrary Poisson—Lie groups.

**Proposition 9.1.** *Consider two $\mathrm{Ad}_G$-invariant functions $f$ and $g$ on a Poisson—Lie group $G$. Then the Poisson bracket $\{f, g\}$ of these functions is also $\mathrm{Ad}_G$-invariant.*

*Proof.* The simplest possible proof of this fact uses Proposition 8.2 in the context of the two-sided action of $G$ on itself. Indeed, the definition of a Poisson—Lie algebra implies that the action of $G \times G^-$ on $G$ itself by

$$(g_1, g_2) \cdot h = g_1 h g_2^{-1}$$

is a Poisson—Lie action. (Here $G^-$ is $G$ endowed with the opposite Poisson structure.) Since the diagonal subgroup $G_\Delta \subset G \times G^-$ is a coisotropic submanifold,[8] the quotient of $G$ by the action of $G_\Delta$ has a natural Poisson structure. This is exactly the claim we are proving.   $\square$

**9.2. Invariant functions on $\widetilde{G}_{\mathrm{int}}$.**   We know already one invariant function on $\widetilde{G}_{\mathrm{int}}$: the degree of a pseudodifferential symbol. However, this function is not very interesting from the point of view of Poisson geometry: it is in involution with any function. On the other hand, this allows us to restrict our attention to any level set of the degree: since the exponential mapping is an isomorphism, the mapping $g \mapsto g^s$ identifies the level sets $t = a$ and $t = sa$, and preserves the adjoint action (it should be mentioned that this mapping is in no *simple* relationship with the Poisson structures on the level sets!). Moreover, any pseudodifferential symbol $L$ can be reduced to be with constant coefficients by the adjoint action of the Lie group of the zero order pseudodifferential symbols with an invertible leading coefficient. Moreover, this reduced form with constant coefficients is uniquely defined:

**Lemma 9.2 ([7]).** *Suppose we work in the case of periodic or rapidly decreasing scalar coefficients. Consider a pseudodifferential symbol $L$ of the order $t \neq 0$ with the leading coefficient 1. Then there exists a pseudodifferential symbol of the form*

$$Q = u_0(x) + u_{-1}(x) D^{-1} + u_{-2}(x) D^{-2} + \dots$$

*with periodic or rapidly decreasing coefficients such that $u_0(x)$ is nowhere 0 and $L_0 = Q \circ L \circ Q^{-1}$ has constant coefficients. The symbol $L_0$ is uniquely determined by $L$.*

*In the matrix case the same is true up to a change that $L$ is generic symbol, and $L_0$ is determined up to a conjugation by a constant matrix.*

---

[8] See the remark after the proposition 8.2.

This lemma can be proved by a simple induction over the (undetermined) coefficients of the symbols $L_0$ and $Q$. It is clear from this proof that the coefficient $u_0$ is determined uniquely up to a multiplicative constant by the subleading coefficient of the symbol $L$.

**Corollary 9.3.** *Any pseudodifferential symbol $L \in \widetilde{G}_{\mathrm{int}}$, $\deg L \neq 0$, can be written in the form $L = Q_0^{-1} f^{-1} L_0 f Q_0$, where the symbol $L_0 \in \widetilde{G}_{\mathrm{int}}$ has constant coefficients, $f$ is an invertible function (i.e., a differential operator of order 0) and $Q_0 \in \widetilde{G}_{\mathrm{int}}$. The symbol $L_0$ is uniquely determined by $L$, the function $f$ is determined up to a multiplicative constant.*

*Proof.* Take $Q = f Q_0$ in the previous lemma. □

Thus we get the following free generators in the ring of invariant functions on $\widetilde{G}_{\mathrm{int}}$: the values of $\frac{f'}{f}$ at different points of $S^1$, and the coefficients of the symbol $L_0$.

**Definition 9.4.** Define the following invariant functions on the group $\widetilde{G}_{\mathrm{int}}$ (in the case of scalar coefficients): For a function $\varphi(x)$ on a circle define $K_\varphi$ as

$$K_\varphi \colon L = \left( 1 + u_{-1}(x) D^{-1} + \dots \right) D^t \mapsto \int \varphi(x) u_{-1}(x) \, dx.$$

Let $H_{-1}$ be the degree $t$ of the symbol $L$, $H_i$ be the coefficient at $D^{-1-i} D^t$ in the symbol $L_0$ from Corollary 9.3.

The next task is to find the commutation relation for these invariant functions (recall that invariant functions form a subalgebra in the Poisson algebra due to Proposition 9.1). As we have already noted, the functions $K_\varphi$ for $\varphi$ with $\int \varphi(x) \, dx = 0$, and the functions $H_i$ are independent and generate the algebra of invariant functions on $\widetilde{G}_{\mathrm{int}}$, moreover, $K_{\varphi \equiv 1} = H_1$. However, the functions $K_\varphi$ can be recognized as characters of 1-dimensional representations, therefore they are Hamiltonians for dressing action of $\widehat{\mathfrak{o}}$. Hence the Hamiltonian flows for these functions (i.e., flows of the corresponding Hamiltonian vector fields) are conjugations by functions. Thus the functions $K_\varphi$ commute as functions in the Heisenberg algebra:

$$\{ K_{\varphi_1}, K_{\varphi_2} \} = \pm H_{-1} \int \varphi_1 \, d\varphi_2.$$

Moreover, the functions $K_\varphi$ commute with $H_i$, since $H_i$ are invariant with respect to the action of $O$ by conjugation. What remains to find out are the commutation relations of $H_i$ and $H_j$.

**Lemma 9.5.** *The functions $H_i$ are in involution.*

We postpone the proof of this lemma until Corollary 9.12.

*Remark* 9.6. The description of the invariant functions $K_\varphi$ generalizes literally to the matrix case. They obviously commute as elements of the affine algebra $\widehat{\mathfrak{gl}_n}$. It is interesting to generalize the functions $H_i$ to the case of matrix coefficients.

**Corollary 9.7.** *The functions $H_i$ can be pushed down to the result of Poisson reduction of $\widetilde{G}_{int}$ with respect to the dressing action of $\widehat{\mathfrak{o}}$. Moreover, these pushes-down are in involution on the reduced manifold.*

*Remark* 9.8. If we consider the result of the Poisson reduction as a submanifold $\bar{\mathcal{M}}^0 = \{u_{-1} \equiv 0\}$ in $\widetilde{G}_{int}$, then these pushes-down are just restrictions of ad-invariant functions to $\widetilde{G}_{int}$. Moreover, $\bar{\mathcal{M}}^0$ is an ad-invariant submanifold (and even a subgroup) of $\widetilde{G}_{int}$, and these functions are exactly ad-invariant functions on $\bar{\mathcal{M}}^0$.

We should note, however, that though the subgroup $\bar{\mathcal{M}}^0$ carries a natural Poisson structure (as a result of Poisson reduction), it *is not* a Poisson—Lie subgroup of $\widetilde{G}_{int}$. In particular, this Poisson structure *is not* compatible with the group structure on $\bar{\mathcal{M}}^0$.

**9.3. The KP Hamiltonians.** Here we investigate the relations of the above Hamiltonians $H_k$ with the wide-known KP Hamiltonians $\widetilde{H}_k$:

$$L = \left(D + u_{-2}(x) D^{-1} + \dots\right) \overset{\widetilde{H}_k}{\mapsto} \operatorname{Tr} L^k, \ k \geq 1.$$

Since Tr is ad-invariant, the functions $\widetilde{H}_k$ are also ad-invariant, thus can be expressed via the functions $H_j$, $j \geq -1$. To find these expressions, we get rid from the restrictions $u_{-1}(x) = 0$ and $\deg L = 1$ by extending the Hamiltonians $\widetilde{H}_k$ to the whole $\widetilde{G}_{int} \smallsetminus G_{int}$ by

$$L = \left(1 + u_{-1}(x) D^{-1} + \dots\right) D^t \overset{\widetilde{H}_k}{\mapsto} \operatorname{Tr} L^{k/t}, \ k \geq 1,$$

$$\widetilde{H}_0 \overset{\text{def}}{=} \operatorname{Tr}\left(\log L - t \log D\right),$$

$$\widetilde{H}_{-1} \overset{\text{def}}{=} H_{-1}.$$

Here we used the good properties of the exponential mapping $\widetilde{\mathfrak{g}}_{int} \to \widetilde{G}_{int}$ to introduce the powers $L^\alpha = \exp \alpha \log L$ of pseudodifferential symbol $L$, the symbol log has the same meaning as in Section 3. Note that $\widetilde{H}_0 = H_0$, and $\log L$ is a natural analogue of $L^\varepsilon$ for an infinitesimally small $\varepsilon$. Since $\deg L^{k/t} = k$, the function Tr is well defined. Note that Tr is not defined for operators of fractional power or for $\log D$.

The above definition is natural analogue of usual Hamiltonians for Gelfand—Dickey structures.

**Theorem 9.9.** *Consider the following system of differential equations on a function of variables $t_k$, $k \geq 1$, with values in $\widetilde{G}_{int}$:*

$$\frac{\partial L}{\partial t_k} = \left[ \left( L^{k/\deg L} \right)_+ , L \right].$$

(1) *These equations are compatible and allow a set of Hamiltonians $\frac{\widetilde{H}_0 \widetilde{H}_k}{k}$ in involution with respect to the second Gelfand—Dickey bracket on $\widetilde{G}_{int}$.*

(2) *These equations preserve the submanifolds $\deg L = $ const;*

(3) *When restricted to the submanifold $\deg L = 1$ they give the (extended) hierarchy of KP equations.*

(4) *After the Poisson reduction by the action of functions they give the usual hierarchy of KP equations.*

(5) *When restricted to the submanifold of differential operators of order $n$ they give the $\mathrm{GL}_n$-KdV hierarchy.*

*Proof.* The proof takes the rest of this section.

**Proposition 9.10.** *Define universal polynomials $\mathcal{H}_k$ of the variables $(t; x_1, x_2, \dots)$ by the identity*

$$\left( 1 + x_1 T + x_2 T^2 + \dots \right)^t \overset{\text{def}}{=} \exp t \log \left( 1 + x_1 T + x_2 T^2 + \dots \right) = 1 + \mathcal{H}_1 T + \mathcal{H}_2 T^2 + \dots$$

*in $\mathbb{Q}[[T]]$. Then $\widetilde{H}_k = \mathcal{H}_{k+1} \left( \frac{k}{H_{-1}}; H_0, H_1, \dots \right)$, $k \geq 1$, $\widetilde{H}_k = H_k$ for $k = 0, -1$. Moreover,*

$$H_k \text{ are polynomials in } \widetilde{H}_k \text{ and } \widetilde{H}_{-1}^{-1}.$$

*Proof.* We need to prove only the last part of the proposition. However, by homogeneity the polynomial $\mathcal{H}_k$ is linear in $t x_k$, therefore $H_{k+1}$ is a polynomial in $\widetilde{H}_{k+1}$ and $H_{-1}^{-1}, H_{-1}, H_0, \dots, H_k$. $\quad \square$

**Example 9.11.**

$$H_{-1} = \widetilde{H}_{-1}, \quad H_0 = \widetilde{H}_0, \quad H_1 = \widetilde{H}_{-1} \widetilde{H}_1 + \frac{1}{2} \widetilde{H}_0^2 - \frac{\widetilde{H}_0^2}{2\widetilde{H}_{-1}},$$

$$H_2 = \frac{1}{2} \widetilde{H}_{-1} \widetilde{H}_2 - \widetilde{H}_0 \widetilde{H}_1 + \frac{1}{2} \widetilde{H}_{-1} \widetilde{H}_0 \widetilde{H}_1 - \frac{1}{12} \widetilde{H}_0^3 + \frac{\widetilde{H}_0^3}{4\widetilde{H}_{-1}} - \frac{\widetilde{H}_0^3}{6\widetilde{H}_{-1}^2}, \quad \dots.$$

**Corollary 9.12.** *The functions from the families $\left\{ \widetilde{H}_k \right\}$ and $\{H_k\}$ are in involution.*

*Proof.* It is sufficient to proof that $\widetilde{H}_k$ are in involution. However, since $\widetilde{H}_{-1} = \deg$, $\widetilde{H}_0 = \phi$, these two functions are Casimir functions. Let us show that the Hamiltonian flow for the function $\widetilde{H}_k$, $k \geq 1$ can be written in a familiar form:

**Lemma 9.13.** *The flow for $\widetilde{H}_k$, $k \geq 1$, is*

$$\frac{dL}{dt_k} = -\frac{k}{t}\left[L, \left(L^{k/t}\right)_+\right] = \frac{k}{t}\left[L, \left(L^{k/t}\right)_-\right]. \qquad t = \deg L \neq 0.$$

*Here the symbol $L^{k/t}$ has an integer order $k$, therefore the operation $()_+$ of taking the differential part makes sense.*

*Proof.* Fix a symbol $L$ of degree $t$. Now the differential of the function $\widetilde{H}_k$ at $L \in \widetilde{G}_{\text{int}}$ can be identified with a symbol of the form $X = D^{-t} \circ \bar{X}$, where $\bar{X}$ is a differential symbol up to addition of integral symbol:

$$\operatorname{Tr}\left(L + \delta L\right)^{k/t} - \operatorname{Tr} L^{k/t} = \operatorname{Tr}\left(D^{-t} \circ \bar{X} \circ \delta L\right).$$

Let us show that we can take $X = \frac{k}{t}L^{\frac{k}{t}-1}$.

If the symbol $L = L_0$ has constant coefficients, then the canonically defined differential part of $\bar{X}$ also has constant coefficients. Thus to find such an $X$ it is sufficient to consider symbols $\delta L$ also with constant coefficients. In this case the left-hand side is $\operatorname{Tr}\left(\frac{k}{t}L_0^{\frac{k}{t}-1} \circ \delta L_0\right)$ up to terms of order $o\left(\delta L\right)$, therefore for symbols with constant coefficients

$$X = \frac{k}{t}L_0^{\frac{k}{t}-1}.$$

If the coefficients of a symbol $L$ are not constant, $L$ can be represented as $QL_0Q^{-1}$, where $L_0$ has constant coefficients, and $Q$ is an invertible pseudodifferential symbol of degree $0$. Since the function $\widetilde{H}_k$ is invariant with respect to conjugation by $Q$, the differential of $\widetilde{H}_k$ at $L$ is the image of the differential at $L_0$ by the action of this conjugation. Therefore in general

$$X = \frac{k}{t}L^{\frac{k}{t}-1}.$$

Now we can apply Formula (6.1) and get

$$\frac{dL}{dt_k} = \frac{k}{t}\left(\left(L^{k/t}\right)_+ L - L\left(L^{k/t}\right)_+\right) = \frac{k}{t}\left[\left(L^{k/t}\right)_+, L\right].$$

$\square$

Moreover, the second formula of Lemma 9.13 shows that the flow for $\widetilde{H}_k$ changes a symbol to a conjugated one, therefore the ad-invariant function $\widetilde{H}_j$ remains constant on a trajectory of the flow. Thus the commutator of $\widetilde{H}_k$ and $\widetilde{H}_j$ vanishes. $\square$

This finishes the proof of the theorem. $\square$

Now we turn back to the relation of two families of Hamiltonians.

*Remark* 9.14. The formulae relating the functions $\widetilde{H}_k$ and $H_k$ show that the functions $\widetilde{H}_k$ have a pole of order $k$ on the hypersurface $\deg L = 0$ in $\widetilde{G}_{\mathrm{int}}$. Therefore it is interesting to consider the flows for $H_k$ instead of flows for $\widetilde{H}_k$. Although Lemma 9.2 is not valid if we drop the restriction $t \neq 0$ (one cannot make the coefficient $u_{-1}(x)$ constant), and the functions $H_k$ have poles on $G_{\mathrm{int}}$ when considered on the whole group $\widetilde{G}_{\mathrm{int}}$, we can fix the Hamiltonians by restricting to a smaller submanifold.

Consider a subgroup $\mathcal{M} \subset \widetilde{G}_{\mathrm{int}}$ consisting of pseudodifferential symbols with a constant subleading coefficient. It is easy to check that Lemma 9.2 remains valid for this subgroup even without the restriction $t \neq 0$. Hence the functions $H_k$ extend smoothly on $\mathcal{M}$. By our definition of functions $H_i$ what we get on $\mathcal{M} \cap \{L \mid \deg L = 0\}$ is the well-known Benney hierarchy of Hamiltonian equations.

*Remark* 9.15. We would like to emphasize here that the above systems preserve $\deg L$, and can be obtained for any $\deg L = t$ by taking the $t$-th power of pseudodifferential symbols from the usual KP hierarchy. Indeed, the operation of taking the $t$-th power identifies level set $\deg L = 1$ with the level set $\deg L = t$ (due to existence of good exp and log), and it is easy to see that the above systems of differential equations are identified under this transformation. Moreover, the Hamiltonian functions are also identified (up to multiplicative constant $t$). However, the *Poisson structures are not identified*, so in fact we get different Hamiltonian realizations of the same hierarchy (cf. [11]).

Thus it is interesting to consider the above systems for a fixed (real or complex) number $t$. As we have seen above, for an integer $t$ this system allows an additional restriction to a submanifold formed by differential operators.

The standard arguments show that these differential equations can be integrated (in the sense that solutions exist for any $t_k$), hence define an infinite set of commuting flows.

*Remark* 9.16. It is a widespread belief that the $n$-KdV systems and KP system (with some natural restrictions) are completely integrable with respect to the second Gelfand—Dickey structure according to any of (equivalent for finite-dimensional systems) definitions of this notion. In particular, this would mean that the result of the Poisson reduction of $\widetilde{G}_{\mathrm{int}}$ with respect to the action of a Lie algebra of invariant functions has a trivial Poisson structure. It is interesting to check this fact and find a general Poisson—Lie-group-theoretic approach to this question.

*Remark* 9.17. Consider a Poisson—Lie subgroup $G_{\mathrm{int}}$ of $\widetilde{G}_{\mathrm{int}}$. We can apply all what we did with $\widetilde{G}_{\mathrm{int}}$ directly to this subgroup. However, in this case we should take the Poisson reduction with respect to much bigger subgroup of the dual group. Indeed, $[\mathfrak{g}_{\mathrm{int}}, \mathfrak{g}_{\mathrm{int}}]$ consists of pseudodifferential symbols of order $\leq -3$, therefore the orthogonal complement consists of differential operators $\mathcal{D}_{\leq 1}$ of order $\leq 1$.

Obviously one can reduce any pseudodifferential symbol of order $-1$ with an invertible leading coefficient to an operator with constant coefficients by a diffeomorphism of the circle and a conjugation by an invertible symbol of order 0. This means that $\mathrm{ad}_{\mathfrak{G}_{\mathrm{int}}}$-*invariant and* $\mathcal{D}_{\leq 1}$-*invariant* functions on this open subset of $G_{\mathrm{int}}$ are exact analogues of the above functions $H_i$ (and coincide with them after the restriction $u_{-1} = \mathrm{const}$). We see that though $H_i$ have a singularity on $G_{\mathrm{int}}$, this singularity "can be compensated" by the fact that the group of diffeomorphisms of $S^1$ acts on $G_{\mathrm{int}}$ and does not act on $\tilde{G}_{\mathrm{int}}$ as on Poisson—Lie groups.

**9.4. Poisson—Lie structure and the exponential mapping.** We want to provide here a different proof of Proposition 9.1, since it uses an interesting fact that seems to be unknown: the relation between the Poisson structure on the Poisson—Lie group and on the corresponding Lie algebra. However, to do this we need to state this relation first.

Consider a Poisson—Lie group $G$ and its Lie bialgebra $(\mathfrak{G}, \mathfrak{G}^*)$. The vector space $\mathfrak{G}$ carries a natural linear Poisson structure, since $\mathfrak{G}$ is identified with a dual space to the Lie algebra $\mathfrak{G}^*$. However, one can also define another Poisson structure on (an open subset of) $\mathfrak{G}$. Indeed, in the points where the exponential mapping is a diffeomorphism we can consider the inverse image of the Poisson structure on $G$. It is easy to see that the former Poisson structure is a linear part of the latter. We are going to characterize the relation between these two Poisson structures more precisely.

For this purpose we recall that a Poisson structure on a manifold $\mathcal{M}$ can be described by a bivector field $\eta$ (i.e., a section of $\Lambda^2 T\mathcal{M}$):

$$\langle \eta|_x, df|_x \wedge dg|_x \rangle = \{f, g\}|_x.$$

Obviously, for any Poisson structure one can find the corresponding bivector field. Moreover, this field uniquely determines this Poisson structure.

**Lemma 9.18.** *Let $\eta^{lin}$ be a bivector field on $\mathfrak{G}$ corresponding to the linear Poisson structure on $\mathfrak{G}$, $\eta^{PL}$ be a bivector field on $\mathfrak{G}$ corresponding to the inverse image of the Poisson structure on $G$ by the exponential mapping. Then*

$$(9.1) \qquad \eta^{PL}|_X = \left( \left( \Lambda^2 \frac{e^{\mathrm{ad}_X} - 1}{\mathrm{ad}_X} \right)^{-1} \circ \frac{e^{\lambda^2 \, \mathrm{ad}_X} - 1}{\lambda^2 \, \mathrm{ad}_X} \right) \cdot \eta^{lin}|_X, \quad X \in \mathfrak{G}.$$

*Here $\mathrm{ad}_X$ maps $\mathfrak{G} \to \mathfrak{G}$, for a linear operator $M \colon V \to V$ we denote by $\lambda^2 M$ and $\Lambda^2 M$ the following mappings $\Lambda^2 V \to \Lambda^2 V$:*

$$\lambda^2 M \colon a \wedge b \mapsto Ma \wedge b + a \wedge Mb, \qquad \Lambda^2 M \colon a \wedge b \mapsto Ma \wedge Mb.$$

*Proof.* We begin with recalling the following simple fact from the theory of Lie groups. Consider an element $X$ in a Lie algebra $\mathfrak{G}$. The differential $d\exp|_X$ of the exponential mapping identifies the tangent space $T_X\mathfrak{G} = \mathfrak{G}$ with $T_{\exp X}G$. On the other hand, we can identify $T_{\exp X} G$ with $\mathfrak{G}$ via left-invariant vector fields, i.e., using the differential $d\mathcal{R}_{\exp X}$ of the right translation on $G$.

**Lemma 9.19.** *These two identifications are related by*

$$d\left(\mathcal{R}_{\exp X}^{-1} \circ \exp\right)|_X = \frac{e^{\operatorname{ad}_X} - 1}{\operatorname{ad}_X}.$$

*Proof.* We provide a proof here since it is useful in the proof of the previous lemma. For a fixed $t$ denote the mapping

$$d\left(\mathcal{R}_{\exp tX_0}^{-1} \circ \exp tX\right)|_{X=X_0} \colon \mathfrak{G} \to \mathfrak{G}$$

by $F(t)$. We express the derivative of the exponential mapping as

$$d|_{X=X_0} \exp tX = d\mathcal{R}_{\exp tX_0} \circ F(t).$$

Now the rule

$$\exp(t + \delta t) X = \exp tX \cdot \exp \delta tX$$

gives us after differentiation

$$d\mathcal{R}_{\exp(t+\delta t)X} \cdot F(t + \delta t) \approx d\mathcal{R}_{\exp \delta tX} \left(d\mathcal{R}_{\exp tX} \cdot F(t)\right) + d\mathcal{L}_{\exp tX} \left(d\mathcal{R}_{\exp \delta tX} \circ F(\delta t)\right),$$

or

$$F(t + \delta t) \approx F(t) + \operatorname{Ad}_{\exp tX} \circ F(\delta t).$$

Thus

$$F'(t) = e^{\operatorname{ad}_{tX_0}} F'(0), \qquad \text{here } ' = \frac{d}{dt}.$$

Evidently $F'(0) = \operatorname{id}$, what proves the lemma. $\qquad \square$

We proceed with the proof of Lemma 9.18 in the same way. The compatibility condition of the multiplication mapping with the Poisson structure can be written as

$$\eta|_{g_1 g_2} = \Lambda^2 d\mathcal{R}_{g_2} \cdot \eta|_{g_1} + \Lambda^2 d\mathcal{L}_{g_1} \cdot \eta|_{g_2}.$$

Here $\mathcal{R}$ and $\mathcal{L}$ denote the mappings of right and left translations on $G$. Denote $\Lambda^2 \mathcal{R}_{\exp -tX} \eta|_{\exp tX}$ as $H(t)$. Now the substitution $g_1 = \exp tX$, $g_2 = \exp \delta tX$ gives us

$$H'(t) = \Lambda^2 e^{t \operatorname{ad}_X} H'(0) = e^{t\Lambda^2 \operatorname{ad}_X} H'(0).$$

And finally $H'(0)$ coincides with $\eta^{\operatorname{lin}}|_X$, what finishes the proof of the lemma. $\qquad \square$

**9.5. A different proof of closeness under Poisson bracket.** Here we provide another proof of Proposition 9.1. This proof is based on Formula (9.1) from Lemma 9.18.

To compute the Poisson bracket of functions on $G$ we can consider their images under the exponential mapping. After this we can apply Formula (9.1):

$$\{f, g\}|_X = \left\langle \eta^{\text{lin}}, \left( \frac{e^{\lambda^2 \operatorname{ad}_X^*} - 1}{\lambda^2 \operatorname{ad}_X^*} \circ \left( \Lambda^2 \frac{e^{\operatorname{ad}_X^*} - 1}{\operatorname{ad}_X^*} \right)^{-1} \right) \cdot df|_X \wedge dg|_X \right\rangle$$

However, if $f$ and $g$ correspond to invariant functions on $G$, then they are $\operatorname{ad}_{\mathfrak{G}}$-invariant functions on the Lie algebra $\mathfrak{G}$, therefore $\operatorname{ad}_X^* \cdot df|_X = 0$, and $\operatorname{ad}_X^* \cdot dg|_X = 0$. Hence the complicated expression we pair with $\eta^{\text{lin}}$ coincides with $df|_X \wedge dg|_X$. Thus the Poisson bracket $\{f, g\}$ of these functions on $G$ goes under the exponential mapping into the Poisson bracket on the Lie algebra $\mathfrak{G}$. Now the only fact we need to prove is the linearized version of Proposition 9.1:

**Lemma 9.20.** *Let $\mathfrak{G}$ be a Lie bialgebra. Consider the corresponding linear Poisson structure on $\mathfrak{G}$. Then the Poisson bracket of two $\operatorname{ad}_{\mathfrak{G}}$-invariant functions on $\mathfrak{G}$ is $\operatorname{ad}_{\mathfrak{G}}$-invariant.*

*Proof.* Note first that on the dual space $\mathfrak{G}^*$ to a Lie algebra an $\operatorname{Ad}_G^*$-invariant function is in involution with *any* function on $\mathfrak{G}^*$. Consider now two covectors $\alpha, \beta \in T_X^* \mathfrak{G}$. These covectors are differentials of $\operatorname{ad}_{\mathfrak{G}}$-invariant functions $f$ and $g$ iff $\operatorname{ad}_X^* \alpha = \operatorname{ad}_X^* \beta = 0$. Consider the multiplication in $\mathfrak{G}^*$ as a mapping $\eta \colon \mathfrak{G} \to \mathfrak{G} \times \mathfrak{G}$. One can express $\{f, g\}|_X$ as

$$\langle \eta|_X, \alpha \wedge \beta \rangle.$$

To show that the Poisson bracket of $f$ and $g$ is $\operatorname{ad}_{\mathfrak{G}}$-invariant, we should show that

$$\left\langle \eta|_{\operatorname{Ad}_g X}, \operatorname{Ad}_g^* \alpha \wedge \operatorname{Ad}_g^* \beta \right\rangle = \langle \eta|_X, \alpha \wedge \beta \rangle,$$

or, infinitesimally, that

$$\left\langle \eta|_{[Y,X]}, \alpha \wedge \beta \right\rangle + \langle \eta|_X, \operatorname{ad}_Y^* \alpha \wedge \beta \rangle + \langle \eta|_X, \alpha \wedge \operatorname{ad}_Y^* \beta \rangle = 0.$$

However, we know that $\eta$ is 1-cocycle of $\mathfrak{G}$ with coefficients in $\mathfrak{G} \otimes \mathfrak{G}$, therefore

$$\left\langle \eta|_{[Y,X]}, \alpha \wedge \beta \right\rangle + \langle \eta|_X, \operatorname{ad}_Y^* \alpha \wedge \beta \rangle + \langle \eta|_X, \alpha \wedge \operatorname{ad}_Y^* \beta \rangle$$
$$- \langle \eta|_Y, \operatorname{ad}_X^* \alpha \wedge \beta \rangle - \langle \eta|_Y, \alpha \wedge \operatorname{ad}_X^* \beta \rangle = 0.$$

Now applying the above conditions on $\alpha$ and $\beta$ we get the required statement. $\qquad \square$

## 10. 4-dimensional extension and the reduced KdV and KP hierarchies

In the preceding section we gave a way to construct the set of (pseudo)differential symbols of the form

$$\left(1 + u_{-1}\left(x\right)D^{-1} + u_{-2}\left(x\right)D^{-2} + u_{-3}\left(x\right)D^{-3} + \dots\right)D^{t}, \qquad u_{-1} = 0,$$

with the second Gelfand—Dickey Poisson bracket as a Poisson submanifold $\{\varphi = 0\}$ of codimension 1 in the quotient (or Poisson reduction) of the Poisson—Lie group $\widetilde{G}_{\mathrm{int}}$ by the dressing action of periodic functions. This Poisson submanifold is a hypersurface and can be included into a codimension 1 foliation of Poisson submanifolds $u_{-1} = \mathrm{const}$.

Note that in this approach we should in fact consider this Poisson manifold as an ordinary member of 1-parameter family. It would be interesting to find out if the members of this family can be identified preserving the Poisson structures.

### 10.1. "Universal" central extension and the Lie bialgebra structure.

In this section we give another construction that results in the following objects:

(1) A (trivial) central extension of the group $\widetilde{G}_{\mathrm{int}}$ together with a Poisson—Lie structure on it (a *Poisson—Lie central extension*);
(2) An extension of the Heisenberg algebra that sits inside the dual group;
(3) A quotient by the action of the extension of the Heisenberg algebra on the extension of $\widetilde{G}_{\mathrm{int}}$.

Moreover, it turns out that this quotient can be naturally identified with the reduced KP hierarchy.

In some way this construction is a little bit "more canonical" than the construction in Section 4. There we defined a central extension, the corresponding dual object (i.e., an extension by $\log D$), and a fusion of these two extensions, i.e., a central extension $\widetilde{\mathfrak{g}}$ of the extension by $\log D$. After this we decomposed the algebra into a direct sum, that gave as a Manin triple.

However, the ambiguity of this construction is in the fact that the Lie algebra of pseudodifferential symbols (say, on $S^1$) has in fact two central extensions: one is given by the cocycle $c_{\log D} = c$ from (2.4), another one by the cocycle $c_x = c^{\circ}$ from (2.5):

$$(10.1) \qquad\qquad c_{\bullet}\left(L, M\right) = \mathrm{Tr}\left(\left[\bullet, L\right] \circ M\right).$$

The 2-cocycle properties are insured by the operations

$$L \mapsto [x, L], \qquad L \mapsto [\log D, L]$$

being outer derivations of the Lie algebra $\mathfrak{g}$: they send (periodic or rapidly decreasing) pseudodifferential symbols into themselves and preserve the Killing form. In fact the cocycle $c_x$ is much more simple than the cocycle $c = c_{\log D}$: it gives a trivial extension of the differential part $\mathfrak{g}_{\mathrm{DO}}$ of $\mathfrak{g}$, in fact, it even vanishes on $\mathfrak{g}_{\mathrm{DO}}$. (Recall that the cocycle $c_{\log D}$ gives the Virasoro extension when restricted to the algebra of vector fields.) Therefore before formulation of Theorem 2.9 we should have chosen a central extension to work with. A different choice of a central extension would result in a different theorem.

We would like to stress here that it is possible to postpone this arbitrariness much longer: we can take the *universal* central extension, and then the arbitrariness is postponed until the choice of decomposition of extended algebra into a sum of two isotropic subalgebras. In particular, *there is a Poisson—Lie structure on any central extension of* $\mathfrak{g}$.

To show this we need the following generalization of Lemma 4.1:

**Lemma 10.1.** Let $\mathfrak{G}$ be a Lie algebra with an ad-invariant bilinear pairing $(,)$. Let $\mathcal{D}$ be a vector space of derivations of $\mathfrak{G}$ that are skew-symmetric with respect to $(,)$. Suppose that a commutator of any two derivations of $D$ is an inner derivation. Consider an arbitrary extension

$$0 \to \mathfrak{G} \to \bar{\mathfrak{G}} \to \mathcal{D} \to 0$$

of $\mathfrak{G}$ corresponding to this "action" of the abelian Lie algebra $\mathcal{D} \to Out\mathfrak{G}$. Fix a splitting $\bar{\mathfrak{G}} = \mathfrak{G} \oplus \mathcal{D}$.

Let $\mathcal{D}^*$ be the dual space to $\mathcal{D}$. Then the vector space

$$\mathfrak{G}_{de} = \bar{\mathfrak{G}} \oplus \mathcal{D}^* = \mathfrak{G} \oplus \mathcal{D} \oplus \mathcal{D}^*$$

carries a natural bracket $[,]_{de}$ with $[,]_{de}$-invariant bilinear form $(,)_{de}$. Here the only non-zero brackets in $\mathfrak{G}_{de}$ are

$$[X_1, X_2]_{de} = [X_1, X_2] + \sum_i \left( d^i(X_1), X_2 \right) C_i,$$

$$[d, X]_{de} = d(X) + \sum_i \left( \left[ d^i, d \right]_{\bar{\mathfrak{G}}}, X \right) C_i, \qquad X, X_1, X_2 \in \mathfrak{G}, \quad d, d_1, d_2 \in \mathcal{D},$$

$$[d_1, d_2]_{de} = [d_1, d_2]_{\bar{\mathfrak{G}}},$$

the pairing remains the same on $\mathfrak{G}$, the subspace $\mathfrak{G}$ is orthogonal to $\mathcal{D} \oplus \mathcal{D}^*$, and $(d^i, d^j)_{de} = (C_i, C_j)_{de} = 0$, $(d^i, C_j)_{de} = \delta^i_j$. In these formulae $(d^i)$ and $(C_i)$ are dual bases of the spaces $\mathcal{D}$ and $\mathcal{D}^*$ correspondingly.

The bracket and the splitting of $\bar{\mathfrak{G}}$ induce the mapping $[,]_{\bar{\mathfrak{G}}} : \mathcal{D} \otimes \mathcal{D} \to \mathfrak{G}$. If the skewsymmetric 4-linear form on $\mathcal{D}$

$$(d_1, d_2, d_3, d_4) \mapsto \sum_{\sigma \in \mathfrak{S}_4} (-1)^\sigma \left( [d_{\sigma_1}, d_{\sigma_2}], [d_{\sigma_3}, d_{\sigma_4}] \right)$$

vanishes, the above bracket on $\mathfrak{G}_{de}$ defines a Lie algebra structure with ad-invariant bilinear form $(,)_{de}$.

The proof of this lemma consists of a direct calculation. In particular, if $\dim \mathcal{D} < 4$, the bracket always satisfies the Jacobi identity. In what follows we need the case $\dim \mathcal{D} = 2$. Note that in the case $\dim \mathcal{D} = 1$ we automatically get an extension of $\mathfrak{G}$ by $\mathcal{D}$ basing only on the outer action. In the case of $\dim \mathcal{D} > 1$ we need to specify the extension separately.

Return to the 2-dimensional central extension of the Lie algebra of differential operators by elements $C_x$ and $C_{\log D}$ (the cocycles $c_x$, $c_{\log D}$ are naturally identified with the coefficients at these elements, i.e., elements of the dual space):

$$0 \to \langle C_x, C_{\log D} \rangle \to \overline{\mathfrak{g}} \to \mathfrak{g} \to 0.$$

Consider a corresponding exact sequence of dual spaces:

$$0 \to \mathfrak{g}^* \to \overline{\mathfrak{g}}^* \to \langle C_x, C_{\log D} \rangle^* \to 0.$$

Here $\mathfrak{g}$ is self-dual, thus one can identify $\mathfrak{g}^*$ with $\mathfrak{g}$ using the Killing form. The above construction gives a natural construction of the space $\langle C_x, C_{\log D} \rangle^*$ and a lifting of this space into $\overline{\mathfrak{g}}^*$.

In particular, beginning with $\mathcal{D} = \langle \mathrm{ad}_x, \mathrm{ad}_{\log D} \rangle$ we obtain the following central extension of the Lie algebra

$$\overline{\mathfrak{g}}^* \stackrel{\mathrm{def}}{=} \left\{ \bar{L} \mid \bar{L} = L + ax + b \log D \right\}, \qquad L \text{ is a pseudodifferential symbol,}$$

by

$$c_x \left( \bar{L}, M \right) = \mathrm{Tr} \left( \left[ x, \bar{L} \right] \circ M \right), \quad \bar{L} \in \overline{\mathfrak{g}}^*,\ M \in \mathfrak{g}; \qquad c_x \left( x, \log D \right) = 0,$$

(these rules uniquely determine $c_x$ by skew-symmetry property), and

$$c_{\log D} \left( \bar{L}, M \right) = \mathrm{Tr} \left( \left[ \log D, \bar{L} \right] \circ M \right), \quad \bar{L} \in \overline{\mathfrak{g}}^*,\ M \in \mathfrak{g}; \qquad c_{\log D} \left( x, \log D \right) = 0.$$

**Theorem 10.2.** *These two 2-cochains are nontrivial independent cocycles for the Lie algebra $\overline{\mathfrak{g}}^*$. The corresponding central extension $\mathfrak{g}^{(2)}$ by elements $C_x$ and $C_{\log D}$ carries a natural $\mathrm{ad}$-invariant inner product defined by the rule $(A, B) \mapsto \mathrm{Tr}\,(A \circ B)$ on the subspace of pseudodifferential symbols, and by*

$$(C_x, x) = (C_{\log D}, \log D) = 1$$

*being the only non-vanishing pairings including the elements $x$, $\log D$, $c_x$, and $c_{\log D}$.*

*Remark* 10.3. It easy to see that the above laws have some strange consequences, like the pseudodifferential symbol 1 ceasing to be in the center of the algebra:

$$[1, x] = -C_{\log D}, \qquad [1, \log D] = C_x.$$

However, they are direct counterparts of Lemma 4.1.

We constructed the Lie algebra $\mathfrak{g}^{(2)}$ in order to determine a Lie bialgebra structure on the 2-dimensional central extension of the Lie algebra of differential operators with periodic coefficients. This Lie bialgebra structure is associated with the structure of a Manin triple, i.e., a decomposition into a direct sum, on the algebra $\mathfrak{g}^{(2)}$. One subspace of this decomposition includes differential operators and the elements $C_x$, $C_{\log D}$, another includes integral pseudodifferential symbols and $x$ and $\log D$, completely analogous to the case of 1-dimensional central extension in Section 4. However, there is another decomposition: the first subspace includes differential operators and $C_{\log D}$, $x$, the second one consists of integral symbols and $C_x$, $\log D$. Similar to the previous considerations, both decompositions determine a Poisson—Lie group structure on the Lie group associated with the second component.[9] (We actually got a one-parameter family of different extensions of $\mathfrak{g}_{\mathrm{DO}}$: the first half always contains subalgebra $\hat{\mathfrak{g}}_{\mathrm{DO}}$ and is contained in the 2-dimensional central extension of $\mathfrak{g}_{\mathrm{DO}} \rtimes \langle x \rangle$. The other half has yet much more degrees of freedom, thus we can provide a lot of different Poisson—Lie structures on any member of this family.) We start with the first decomposition.

**10.2. The first decomposition.**   The corresponding extension $\bar{G}_{\mathrm{int}}$ of the Lie group of pseudodifferential symbols of negative order is generated by $D^t$, $e^{sx}$ and $(1 + L_{-1})$, $\deg L_{-1} \leq -1$. Therefore the generic element of this group is

$$(10.2) \qquad\qquad e^{sx} \left(1 + L_{-1}\right) D^t.$$

The center of the dual Lie algebra is 3-dimensional and contains elements $c_x$, $c_{\log D}$ and 1. To describe Casimir functions on the group we note that the corresponding left-and-right-invariant 1-forms on the Lie group $\bar{G}_{\mathrm{int}}$ are $ds$, $dt$ and $d\phi - t\,ds$ (here $\phi$ is the mean value of the leading coefficient $u_{-1}$ of the symbol $L_{-1}$). Therefore the corresponding Casimir functions with left-invariant differentials are $s$ and $t$ from (10.2). However, the third 1-form can be corrected by addition of the product of the first by the antidifferential of the second to get $d\phi$. This gives us three Casimir functions $s$, $t$, and $\phi$ on the group.

*Remark* 10.4. We do not know any algebraic explanation of this anomaly, when we should correct one left (and right) invariant 1-form by some combination of others to get a closed form, i.e., a Casimir function.

Consider now the Poisson structure on the group $\bar{G}_{\mathrm{int}}$ and its relation to the Poisson structure on the group $\widetilde{G}_{\mathrm{int}}$ investigated above. It is easy to see that $\widetilde{G}_{\mathrm{int}}$ is a Poisson subspace of $\bar{G}_{\mathrm{int}}$, because this is true for corresponding Poisson—Lie algebras. Moreover, the vector space generated by $x$ is also a Poisson—Lie subalgebra, hence the action by (say) left multiplication preserves a Poisson structure.

---

[9]The first component cannot be integrated to a group.

Thus as a Poisson manifold $\overline{G}_{\text{int}}$ is a direct product of $\widetilde{G}_{\text{int}}$ and the 1-parametric subgroup $\{e^{sx}\}$.

However, as a group it is only a semidirect product, therefore we should begin our search for invariant functions from scratch. It turns out that on $\bar{G}_{\text{int}}$ *there is no invariant functions at all* but the functions of $t$ and $s$. Indeed, for a generic pair $(s, t)$ the action of the Lie algebra of integral symbols spans the whole tangent space.

Consider now the (left) dressing action. Here we show that there is no essential difference between the quotients of $\widetilde{G}_{\text{int}}$ and $\bar{G}_{\text{int}}$ by the dressing action of the algebra of functions.

Indeed, this action obviously goes via the non-extended Lie algebra of differential operators (that is the same for both cases). The action on the Poisson submanifold $\widetilde{G}_{\text{int}}$ is the same. Now the action of $\{e^{sx}\}$ by left multiplication preserves both the Poisson structure and left-invariant 1-forms, therefore it commutes with the dressing action. Therefore the left dressing action on $s = s_0$ is isomorphic to the left dressing action on $s = 0$.

In the same way if we consider the translation of the right dressing action from $s = s_0$ to $s = 0$ by left action of $\{e^{sx}\}$, it differs from the right dressing action on $s = 0$ by conjugation by $e^{s_0x}$ on $s = 0$, which preserves the Poisson structure. This conjugation preserves the decomposition of $\widetilde{\mathfrak{g}}$ into $\widetilde{\mathfrak{g}}_{\text{int}} \oplus \widehat{\mathfrak{g}}_{\text{DO}}$ and preserves the invariant form on $\widetilde{\mathfrak{g}}$. This allows us to conclude that instead of this conjugation on $\widetilde{G}_{\text{int}}$ one can make a conjugation on $\widehat{\mathfrak{g}}_{\text{DO}}$: the action of $X$ on $e^{s_0x}L_0e^{-s_0x}$ is the same as the action of $e^{-s_0x}Xe^{s_0x}$ on $L_0$ (i.e., the second tangent vector field goes to the first by the mapping $L \mapsto e^{s_0x}Le^{-s_0x}$). We see that there is no essential difference between the quotients of $\widetilde{G}_{\text{int}}$ and $\overline{G}_{\text{int}}$ by the dressing action of the algebra of functions, the only one is the absence of invariant functions away from special submanifolds.

## 10.3. The second decomposition.

Consider now the second decomposition into a sum of two isotropic subalgebras. To simplify the notations we continue to denote these subspaces by $\overline{\mathfrak{g}}_{\text{int}}, \bar{\mathfrak{g}}_{\text{DO}}$, though they differ from the subspaces in the last paragraphs.

Since $C_x$ generates a Poisson—Lie subalgebra in $\overline{\mathfrak{g}}_{\text{int}}$, the left multiplication by the corresponding 1-parametric subgroup preserves the Poisson structure on $\overline{G}_{\text{int}}$. Moreover, this subgroup is in the center of $\overline{G}_{\text{int}}$, thus as a Lie group $\overline{G}_{\text{int}}$ is a direct product $\widetilde{G}_{\text{int}} \times \mathbb{R}$. Therefore, except for the coefficient at $C_x$ in $e^{sC_x}$, the invariant functions on $\overline{G}_{\text{int}}$ are the same as on $\widetilde{G}_{\text{int}}$ (and, in particular, all the KP-Hamiltonians are invariant on $\overline{G}_{\text{int}}$).

However, $\overline{G}_{\text{int}}$ is not a direct product as a Poisson manifold. Indeed, $\widetilde{G}_{\text{int}}$ is not a Poisson submanifold, since $\widetilde{\mathfrak{g}}_{\text{int}}$ is not a Poisson—Lie subalgebra. The difference between the Poisson structure on $\overline{G}_{\text{int}}$ and the Poisson structure on $\widetilde{G}_{\text{int}} \times \mathbb{R}$ is a

2-vector field

$$-\frac{\partial}{\partial s} \wedge \frac{d}{d\xi}$$

on $\overline{G}_{\text{int}}$. Here $s$ is a parameter on $\left\{e^{sC_x}\right\}$, $\frac{d}{d\xi}$ is a vector field on $\widetilde{G}_{\text{int}}$ given by the formula

$$L \mapsto L - \varepsilon\,[x, L]$$

with an infinitesimal parameter $\varepsilon$.

The center of $\overline{\mathfrak{g}}_{\text{DO}}$ is 1-dimensional and is generated by $C_{\log D}$. The corresponding Casimir function on $\overline{G}_{\text{int}}$ is $t$.

Consider now the dressing action by the algebra of functions. In this case we consider not only the central extension of this algebra by the element $C_{\log D}$, but also the extension by the element $x$. Now the element 1 of this algebra ceases to be in the center, since

$$[1, x] = -C_{\log D}.$$

This algebra continues to act by conjugation, as in the case of $\widetilde{G}_{\text{int}}$. It is clear that the dressing action of periodic functions is the same as in the case of $\widetilde{G}_{\text{int}}$, the element $x$ acts by $-\frac{d}{d\xi}$, and the center acts trivially. We see that the Hamiltonians for the action are values of the leading coefficient $u_{-1}$ of $L_{-1}$ in the representation $e^{sC_x}\left(1 + L_{-1}\right)D^t$ at different points of the line:

$$H_f = \int_0^1 f\,(x)\,u_{-1}\,(x)\,dx, \quad H_{C_{\log D}} = t, \quad H_x = s.$$

The Poisson reduction coincides again with the quotient. Since a choice of the point in the coalgebra fixes $t$, $s$ and $u_{-1}\,(x)$, we can see that the Poisson reduction corresponds to a choice of values of this parameters. Different values of $u_{-1}$ and $s$ correspond to a choice of different points on a coadjoint orbit if $t \neq 0$, therefore they give the same Poisson reduction.

Thus for a fixed $t \neq 0$ the Poisson reduction of $\overline{G}_{\text{int}}$ by the dressing action of the algebra of functions is exactly the reduced KP Poisson structure, and the KP Hamiltonians are exactly the images of the invariant functions on the Poisson—Lie group.

The reduced $n$-KdV hierarchy can be obtained in the same way taking a Poisson submanifold in the hypersurface $t = n \in \mathbb{N}$.

**10.4. The Fourier transform.** Let us return for a moment to the non-extended Lie algebra of pseudodifferential symbols. Throughout this paper we considered pseudodifferential symbols with smooth (periodic) coefficients $u_m\,(x)$. Consider now the coordinate change on the circle $S^1$: $x = i\log z$, i.e., consider this circle as a unit circle in the complex $z$-plane. Now instead of functions on $S^1$ we can consider, say, functions in the unit disk, or, better, functions on an infinitesimally small circle, i.e., the algebra $\mathbb{C}\left((1/z)\right)$ of jets of rational functions at $z = \infty$ (this choice is better than $z = 0$, as shows the following discussion).

Moreover, since $\frac{d}{dx} = iz\frac{d}{dz}$, any pseudodifferential symbol on $S^1$ can be written as a pseudodifferential symbol in $z$ with coefficients in functions on $\{|z| = 1\}$. The composition law of pseudodifferential symbols in this basis is given by the same Formula (2.1) with the only changes of $x$ to $z$, and of $\xi$ to $\zeta$, where $\zeta$ represents $\frac{d}{dz}$. Thus instead of going via a new coordinate $z$ we could as well consider pseudodifferential symbols with coefficients in $\mathbb{C}\left((x^{-1})\right)$ from the beginning. However, in the above approach it is more clear that the formula for the Tr can be extended to this case. Now it can be written in a symmetrical form:

$$\operatorname{Tr} L\left(z, \zeta\right) = \operatorname*{res}_{z} \operatorname*{res}_{\zeta} L\left(z, \zeta\right).$$

This formula demonstrates the new feature of the new algebra we consider: it has an additional symmetry between $z$ and $\zeta$. Indeed, an element of the algebra is a Laurent series in $\zeta^{-1}$ that has only a finite number of terms with positive powers of $\zeta$, and a coefficient at $\zeta^m$ is a Laurent series in $z^{-1}$ that has only a finite number of terms with positive powers of $z$. To achieve a complete symmetry we can demand the orders of poles at $z = \infty$ of coefficients $u_m(z)$ to be bounded from above.

This algebra allows a remarkable automorphism of order 4: the Fourier transform, which sends $z$ to $i\zeta$, and $\zeta$ to $iz$. It is easy to see that the Fourier transform determines an involution in the $H_2(\mathfrak{g})$, which interchanges $C_{\log D}$ and $C_x = C_{\log z}$. Thus this involution can be extended to the Lie algebras $\bar{\mathfrak{g}}$, $\bar{\mathfrak{g}}^*$ and $\mathfrak{g}^{(2)}$.

The Fourier-transformed constructions of Manin triples define Lie bialgebra structures on the Lie algebras of pseudodifferential symbols of different kinds with analytic coefficients at $z = \infty$. It would be very interesting to give an independent description of the Poisson—Lie groups corresponding to these Lie bialgebras.

## 11. Results for Hopf algebras

This section plays a separate rôle: we allow ourselves to be vague and succinct and discuss some unpolished ideas and open questions. Here we provide the interpretation of several results of this paper from the viewpoint of Hopf algebras. Recall that a Poisson—Lie group is the 1-term expansion at $h = 0$ of a deformation of a commutative Hopf algebra. Thus we can, for example, conjecture that there exists a Hopf algebra with a parameter $h$ such that at $h = 0$ we get the group $\tilde{G}_{\text{int}}$, and the linear part of this deformation corresponds to the Poisson structure on $\tilde{G}_{\text{int}}$. After this we can pose the questions for which of results obtained here there exists a generalization to an arbitrary $h$.

**11.1. KP Hamiltonians.** First of all, our description of KP-hamiltonians allows translation for an arbitrary Hopf algebra in place of the ring of functions on $G$. Indeed, this Hamiltonians are, first of all, ad-invariant functions on the group. However, in the case of Hopf algebras there are two possibilities. The corresponding to ad-action notion for a Hopf algebra $H$ is the left (or right) adjoint action of the dual Hopf algebra $H^*$ on $H$. The invariant functions with respect to such an action (i.e., functions on which $H^*$ acts as a counit $\varepsilon$) form an algebra.

Second, the KP-Hamiltonians form a center of the Lie algebra of ad-invariant functions with respect to the Poisson bracket. The corresponding notion for an arbitrary Hopf algebra is the notion of center of associative algebra (with respect to the $*_h$-product in the case of deformation).

**Definition 11.1.** A (left) KP-Hamiltonian in a Hopf algebra $H$ is an element of the center of the associative subalgebra of functions that are invariant with respect to the (left) adjoint action:

$$\mathcal{KP}\left(H\right) = \operatorname{Cent} H^{H^*_{\mathrm{ad}^{(L)}}}, \quad H^{H^*_{\mathrm{ad}^{(L)}}} = \left\{ f \in H \mid \mathrm{ad}_g^{(L)} f = \varepsilon\left(g\right) f \quad \forall g \in H^* \right\}.$$

Her $\varepsilon$ is counit for $H^*$, $\mathrm{ad}^{(L)}$ is the mentioned above left adjoint action.

To explain the second formula in the definition and the relation to the formula in the Poisson—Lie case

$$\mathcal{KP} = \operatorname{Cent} \operatorname{Func}\left(\operatorname{Group}\right)^{\mathrm{Ad}_{\mathrm{Group}}}$$

recall that for an action of associative algebra $\mathcal{A}$ with a counit $\varepsilon$ (say, for a universal enveloping algebra $U\left(\mathfrak{G}\right)$) the set of fixed vectors in a representation $V$ can be written as

$$V^{\mathcal{A}} = \left\{ v \in V \mid av = \varepsilon\left(a\right) v \quad \forall a \in \mathcal{A} \right\}.$$

**Conjecture 11.2.** For an appropriate quantization $\mathcal{H}_t$ of $\widetilde{G}_{\mathrm{int}}$ the limit of $\mathcal{KP}\left(\mathcal{H}_t\right)$, $t \to 0$, coincides with the algebra generated by KP Hamiltonians on $\widetilde{G}_{\mathrm{int}}$.

This would mean that any KP Hamiltonian survives after quantization.

**11.2. $W_\infty$ and the second KP Poisson structure.** Suppose again that we could quantize the group $\widetilde{G}_{\mathrm{int}}$. Denote by $\mathcal{H}_h$ the corresponding Hopf algebra, $\mathcal{H}_0$ being the Hopf algebra of functions on $\widetilde{G}_{\mathrm{int}}$. Since we cannot provide such a quantization anyway, we allow ourselves to be not very strict below.

It is a widespread belief that so called $W_\infty$ algebra (i.e., the modified in one or another way Lie algebra of differential operators) is closely related to the KP Poisson structure. Here we show they are indeed related, and the relation is like the relation between two limits with order of variables going to limit interchanged. Here for simplicity we work with the extended KP manifold

$$\left\{ L = D + u_0\left(x\right) + u_{-1}\left(x\right)D^{-1} + u_{-2}\left(x\right)D^{-2} + \dots \right\}.$$

First we want to give a verbal description of what we want to do. Consider first the nonquantized variant. The exponential mapping $\widetilde{\mathfrak{g}}_{\mathrm{int}} \to \widetilde{G}_{\mathrm{int}}$ identifies the Poisson structures on these two Poisson manifolds (recall that $\widetilde{\mathfrak{g}}_{\mathrm{int}} = \widehat{\mathfrak{g}}_{\mathrm{DO}}^*$, thus carries a Poisson structure) up to a linear term in the origin. The hypersurface $\{\deg L = \mathrm{const}\}$ corresponds under this identification to the hyperplane $c = \mathrm{const}$ in $\widehat{\mathfrak{g}}_{\mathrm{DO}}^*$ (here $c$ is the central charge for $\widehat{\mathfrak{g}}_{\mathrm{DO}}$). We see that the Poisson algebra of functions on $\{\alpha \in \widehat{\mathfrak{g}}_{\mathrm{DO}}^* \mid c\left(\alpha\right) = \mathrm{const}\}$ is a good approximation to the Poisson algebra of functions on $\left\{ L \in \widetilde{G}_{\mathrm{int}} \mid \deg L = \mathrm{const} \right\}$ if const is sufficiently small. (In fact we should consider not only small $\deg L$, but also operators $L$ with "small" coefficients, for $L$ to be a small neighborhood of the unity.)

On the other hand, (polynomial) functions on $\widehat{\mathfrak{g}}_{\mathrm{DO}}^*$ can be described as $S^\bullet\left(\widehat{\mathfrak{g}}_{\mathrm{DO}}\right)$, and the Poisson bracket on this Poisson algebra is the natural Lie algebra structure on $S^\bullet\left(\widehat{\mathfrak{g}}_{\mathrm{DO}}\right)$. Hence we can identify the functions on $\{\alpha \in \widehat{\mathfrak{g}}_{\mathrm{DO}}^* \mid c\left(\alpha\right) = \mathrm{const}\}$ with the quotient Poisson algebra $S^\bullet\left(\widehat{\mathfrak{g}}_{\mathrm{DO}}\right)/\left(c = \mathrm{const}\right)$. Slightly abusing the standard notations, we can denote the Poisson algebra of functions on $\left\{ L \in \widetilde{G}_{\mathrm{int}} \mid \deg L = \delta \right\}$ as $\widetilde{W}_\delta$. (The "genuine" $W$-algebras are defined only for natural $\delta = k$ and are not functions on $\left\{ L \in \widetilde{G}_{\mathrm{int}} \mid \deg L = k \right\}$, but on a Poisson submanifold $\mathcal{D}_k$ of purely differential operators.)

Thus the above relation can be written as

$$S^\bullet\left(\widehat{\mathfrak{g}}_{\mathrm{DO}}\right)/\left(c = \delta\right) \overset{\delta \to 0}{=} \widetilde{W}_\delta.$$

However, the Poisson algebra $S^\bullet\left(\mathfrak{G}\right)$ is not very natural object to study. The only place where it appears naturally is the semiclassical limit of the associative algebra $U\left(\mathfrak{G}\right)$. It is natural to conjecture that since the left-hand side has a natural quantization $U\left(\widehat{\mathfrak{g}}_{\mathrm{DO}}\right)/\left(c = \delta\right)$, the right-hand side should also have such a quantization that preserves the above approximation. Let us investigate such a quantization, or, better, a quantization of the whole $\widetilde{G}_{\mathrm{int}}$ as above.

Consider the function $\deg$ on $\widetilde{G}_{\mathrm{int}}$. The extended KP manifold is the level set $\deg L = 1$. Thus the algebra of functions on it is

$$\widetilde{W}_1 = \mathcal{H}_0/\left(\deg -1\right)\mathcal{H}_0.$$

We saw in Section 6 that the function $\deg$ is the Casimir function with respect to the Poisson structure. Hence it remains in the center of $\mathcal{H}_h$ up to the first term in expansion in $h$. It is reasonable to conjecture that it remains in the center for any $h$. Since we need this conjecture to define quantizations of algebras $\widetilde{W}_\delta$, let us conjecture it. Then the quantization of $\widetilde{W}_\delta$ is $\mathcal{H}_h / (\deg = \delta)$, in particular, quantization of KP structure is the associative algebra $\mathcal{H}_h / (\deg = 1)$.

On the other hand, consider an arbitrary Hopf algebra $\mathbb{H}_h$ with parameter $h$. Suppose that $\mathbb{H}_0 = \mathrm{Func}\,(G)$, and the first term in $h$ near $h = 0$ of this deformation corresponds to some Poisson—Lie structure on $G$ with corresponding Lie bialgebra $(\mathfrak{G}, \mathfrak{G}^*)$. Consider the linearization of this Poisson structure near the unity. It clearly coincides with the Lie—Beresin—Kirillov—Kostant structure on the dual space $\mathfrak{G}$ to the Lie algebra $\mathfrak{G}^*$.

However, the linearization itself is a result of a limit process, when we consider the manifold in question in bigger and bigger magnification. Thus we get the structure of bialgebra as a result of two subsequent limit processes: one corresponds to $h \to 0$, another one to the characteristic length unit on the group $G$ going to 0. However, we can get the same Poisson structure as a limit when only $h$ goes to 0, but for this we should start from a *different* family of associative algebras $U\left(\mathfrak{G}^*_{(h)}\right)$. Here for any Lie algebra $\mathfrak{G}$ we denote by $\mathfrak{G}_{(h)}$ a Lie algebra that is isomorphic to $\mathfrak{G}$ as a vector space and carries a commutator

$$[X, Y]_{\mathfrak{G}_{(h)}} = h\,[X, Y]_{\mathfrak{G}}\,.$$

(It is evidently just a rescaling of $\mathfrak{G}$ in $h$ times if $h \neq 0$.) Indeed, identify a polynomial on $\mathfrak{G}_{(h)} = \mathfrak{G}$, i.e., an element of $S^\bullet \mathfrak{G}_{(h)} = S^\bullet \mathfrak{G}$, with an element of $U\left(\mathfrak{G}_{(h)}\right)$ by symmetrization. Now if we could interchange two limits in the right-hand side of

$$\lim_{h \to 0} U\left(\mathfrak{G}_{(h)}\right) = \lim_{\mathrm{scale} \to 0} \lim_{h \to 0} \mathcal{H}_h,$$

then we could "identify" $U\left(\mathfrak{G}_{(h)}\right)$ with a small-scale limit (=linearization) of $\mathcal{H}_h$. In other words, $\mathcal{H}_h$ would be a non-linear variant of $U\left(\mathfrak{G}_{(h)}\right) \simeq U\left(\mathfrak{G}\right)$.

Let us apply this discussion to the case $G = \widetilde{G}_{\mathrm{int}}$. We can describe the universal enveloping algebra of centrally extended differential operators as a "linearization" of the would-be deformation of quantization of the group $\widetilde{G}_{\mathrm{int}}$ (up to exchange of limits order). When we reduce the scale the hypersurface $\deg = 1$ goes further and further from the origin. Since the element of $\widehat{\mathfrak{g}}_{\mathrm{DO}}$ that corresponds to the linearization of $\deg$ is the central element $C \in \widehat{\mathfrak{g}}_{\mathrm{DO}}$, than the best approximation to $\deg = 1$ we can choose is $c = \infty$. Thus

$$U\left(\widehat{\mathfrak{g}}_{\mathrm{DO}}\right) / (c = \infty) \text{ is a linearization of quantum KP manifold}$$

(up to an interchange of two limits).

We can make this argument a little bit more precise by introducing one more parameter $\delta$ and considering a limit of

$$\mathcal{H}_h / (\deg = \delta)$$

when $\delta \to 0$. In the classical picture ($h = 0$) this corresponds to the consideration of pseudodifferential operators of a *very small* order. This hypersurface contains a part that is near to the unit element of the group, thus allows a linearization. The above identity becomes

$$\lim_{h \to 0} \left( U \left( \mathfrak{G}_{(h)} \right) / c = f(h) \right) = \lim_{\text{scale} \to 0} \lim_{\delta \to 0} \lim_{h \to 0} \mathcal{H}_h / \left( \deg = \delta \right),$$

here we need to introduce an unknown function $f(h)$ that describes the relation of the central charge with the expansion parameter $h$ on the left-hand side. While the former formula did not allow a precise variant, with the latter formula we can expect that the associative algebra $U(\mathfrak{G}) / (c = \text{const})$ can be obtained as a honest limit of the right hand side when the scale and the parameters $\delta$ and $\varepsilon$ go to a limit with some particular relations between them. Thus we come to

**Conjecture 11.3.** For an appropriate quantization $\mathcal{H}_h$ of $\widetilde{G}_{\text{int}}$ we can realize $U(\widehat{\mathfrak{g}}_{\text{DO}}) / (c = \gamma)$ as a limit

$$U(\widehat{\mathfrak{g}}_{\text{DO}}) / (c = \gamma) = \lim_{\text{scale}, \delta, h \to 0} \mathcal{H}_h / (\deg = \delta),$$

here $\gamma$ is an arbitrary number, in the limit process the variables decrease along an appropriate trajectory that depends on $\gamma$. Here "scale" under the limit sign means that we identify different algebras $\mathcal{H}_h / (\deg = \delta)$ using some analogue of rescaling in a non-local situation.

What physicists did was to consider a limit $\deg \to \infty$ in the Poisson—Lie group $\widetilde{G}_{\text{int}}$ (they also took restrictions on the submanifolds of differential operators of growing size, what was probably irrelevant). In our description we consider a hypersurface in a very small neighborhood of unity (hence of small deg) instead, but since the Plank constant $h$ decreases, we should rescale our picture to preserve the commutation relations (this is the standard process of linearization). Thus in the rescaled coordinates the hypersurface goes to infinity (if we preserve the degree), i.e., the effective central charge we consider increases. In other words, we consider very small central charges for a family of algebras, but if we consider approximate isomorphisms with a fixed algebra, the central charge in the latter algebra goes to infinity. This is the relation of our description with the physical one.

One of the most intriguing open questions about this Poisson—Lie geometry of pseudo-differential operators is a construction of a Drinfeld—Sokolov reduction. The actual Drinfeld—Sokolov reduction gives the Poisson structures on differential operators of an integer order as a reduction of an appropriate affine Lie algebra. The question is to find such a reduction procedure for the whole group of arbitrary order pseudodifferential symbols simultaneously.

DEPARTMENT OF MATHEMATICS, YALE UNIVERSITY, NEW HAVEN, CT 06520
*E-mail address*: khesin@math.yale.edu

DEPARTMENT OF MATHEMATICS, MIT, CAMBRIDGE, MA 02139
*E-mail address*: ilya@math.mit.edu